\documentclass[prd,twocolumn,showpacs,preprintnumbers,amsmath,amssymb,nofootinbib]{revtex4}
\usepackage{graphicx}% Include figure files
\usepackage{dcolumn}% Align table columns on decimal point
\usepackage{bm}% bold math
\usepackage{color}
\usepackage{units}
\usepackage{amsmath}

\usepackage[caption=false]{subfig}
		
\usepackage[colorlinks=true,
	linkcolor=black,
	citecolor=black,
	urlcolor=black,
	filecolor=black
]{hyperref}

\newcommand{\eq}[1]{Eq.~(\ref{#1})}

\newcommand{\bdtau}{\ensuremath{B\to D\tau\nu}}
\newcommand{\bdstau}{\ensuremath{B\to D^*\tau\nu}}
\newcommand{\btau}{\ensuremath{B\to \tau\nu}}
\newcommand{\rd}{\ensuremath{{\cal R}(D)}}
\newcommand{\rds}{\ensuremath{{\cal R}(D^*)}}
\newcommand{\rdrds}{\ensuremath{{\cal R}(D^{(*)})}}

\newcommand{\matrixx}[1]{\begin{pmatrix} #1 \end{pmatrix}} %Matrix with brackets

\newcommand{\hc}{\mathrm{h.c.}}

\newcommand{\dd}{\mathrm{d}}

\allowdisplaybreaks

\begin{document}
\preprint{\vbox{\hbox{CERN-PH-TH-2015-168}\hbox{ULB-15-13}}}

\title{A perturbed lepton-specific two-Higgs-doublet model facing experimental hints for physics beyond the Standard Model}
\author{Andreas Crivellin}
\email{andreas.crivellin@cern.ch}
\affiliation{CERN Theory Division, CH--1211 Geneva 23, Switzerland}%
\author{Julian Heeck}
\email{julian.heeck@ulb.ac.be}
\affiliation{Service de Physique Th\'eorique, Universit\'e Libre de Bruxelles, Boulevard du Triomphe, CP225, 1050 Brussels, Belgium}
\author{Peter Stoffer}
\email{stoffer@hiskp.uni-bonn.de}
\affiliation{Helmholtz-Institut f\"ur Strahlen- und Kernphysik (Theory) and Bethe Center for Theoretical Physics, University of Bonn, D--53115 Bonn, Germany}%

\begin{abstract}
The BaBar, Belle, and LHCb collaborations have reported evidence for new physics in \bdtau{} and \bdstau{} of approximately $3.8\sigma$. There is also the long lasting discrepancy of about $3\sigma$ in the anomalous magnetic moment of the muon, and the branching ratio for $\tau\to\mu\nu\nu$ is $1.8\sigma$ ($2.4\sigma$) above the Standard Model expectation using the HFAG (PDG) values. Furthermore, CMS found hints for a non-zero decay rate of $h\to\mu\tau$. Interestingly, all these observations can be explained by introducing new scalars. In this article we consider these processes within a lepton-specific two-Higgs doublet model (i.e.~of type X) with additional non-standard Yukawa couplings. It is found that one can accommodate $\tau\to\mu\nu\nu$ with modified Higgs--$\tau$ couplings. The anomalous magnetic moment of the muon can be explained if the additional neutral CP-even Higgs $H$ is light (below 100 GeV). Also \rd{} and \rds{} can be easily explained by additional $t$--$c$--Higgs couplings. Combining these $t$--$c$ couplings with a light $H$ the decay rate for $t\to H c$ can be in a testable range for the LHC. Effects in $h\to\mu\tau$ are also possible, but in this case a simultaneous explanation of the anomalous magnetic moment of the muon is difficult due to the unavoidable $\tau\to\mu\gamma$ decay.
\end{abstract}
\maketitle

\section{\label{sec:level1}Introduction}

In addition to direct searches for new physics (NP) performed at very high energies at the LHC, low-energy precision flavour observables provide a complementary window to physics beyond the Standard Model (SM). Here, the anomalous magnetic moment of the muon $(g-2)_\mu$ is a prime example as it is very sensitive to physics beyond the SM entering via quantum corrections. Also, tauonic $B$ meson decays and $\tau\to\mu\nu\nu$ are excellent probes of NP: they test lepton flavour universality -- satisfied in the SM -- and are sensitive to new degrees of freedom that couple proportional to the mass of the involved particles (e.g.\ Higgs bosons~\cite{Krawczyk:1987zj}) because of the heavy $\tau$ lepton involved. The observation of flavour violating decays of the SM Higgs, most importantly $h\to\mu\tau$, would prove the existence of physics beyond the SM.

Let us briefly review the current experimental and theoretical situation in these processes. The world average of the measurement of $a_\mu \equiv (g-2)_\mu/2$ is completely dominated by the Brookhaven experiment E821~\cite{Bennett2006} and is given by~\cite{Olive2014}
\begin{align}
	a_\mu^\mathrm{exp} = (116\,592\,091\pm54\pm33) \times 10^{-11} \,,
\end{align}
where the first error is statistical and the second systematic. The current SM prediction is
\begin{align}
	a_\mu^\mathrm{SM} = (116\,591\,855\pm59) \times 10^{-11} \,,
\end{align}
where almost the whole uncertainty is due to hadronic effects.\footnote{The SM prediction includes the QED corrections~\cite{Aoyama2012}, an electro-weak contribution~\cite{Czarnecki1995,Czarnecki1996,Gnendiger2013}, leading order, next-to-leading order~\cite{Davier2011,Hagiwara2011} and next-to-next-to leading order~\cite{Kurz2014} hadronic vacuum polarisation contributions and the model-dependent leading order~\cite{Jegerlehner2009} and next-to leading order~\cite{Colangelo2014} hadronic light-by-light contributions.} This amounts to a discrepancy between the SM and experimental value of $a_\mu^\mathrm{exp}-a_\mu^\mathrm{SM} = (236\pm 87)\times 10^{-11}$, i.e.~a $2.7\sigma$ deviation.
It is not yet clear if this discrepancy is due to NP or rather underestimated hadronic uncertainties; there are ongoing efforts to reduce the model dependence in the hadronic light-by-light estimate based on dispersion relations~\cite{Colangelo2014a,Colangelo2014b,Colangelo2015} or lattice QCD~\cite{Hayakawa2006,Blum2012,Blum2015,Green2015}. Possible NP explanations besides supersymmetry (see for example Ref.~\cite{Stockinger:2006zn} for a review) include very light $Z^\prime$ bosons~\cite{Langacker:2008yv,Baek:2001kca,Ma:2001md,Gninenko:2001hx,Pospelov:2008zw,Heeck:2011wj,Harigaya:2013twa,Altmannshofer:2014pba}, lepto\-quarks~\cite{Chakraverty:2001yg,Cheung:2001ip}, additional fermions \cite{Freitas:2014pua} but also new scalar contributions in two-Higgs-doublet models (2HDM)~\cite{Iltan:2001nk,Omura:2015nja}, also within the lepton-specific 2HDM~\cite{Broggio:2014mna,Wang:2014sda,Abe:2015oca}.

Concerning tauonic $B$ decays, the BaBar collaboration performed an analysis of the semileptonic $B$ decays \bdtau{} and \bdstau{} using the full available data set~\cite{BaBar:2012xj}. Recently, these decays have also been reanalyzed by Belle~\cite{BDstaumuBELLE,Huschle:2015rga}, and LHCb measured the mode \bdstau{}~\cite{Aaij:2015yra}. These experiments find for the ratios
\begin{align}
	{\cal R}(D^{(*)})\,\equiv\,\frac{{\rm BR}(B\to D^{(*)} \tau \nu)}{{\rm BR}(B\to D^{(*)} \ell \nu)} \,,
\end{align}
where $\ell = e$ or $\mu$, the following values:
\begin{align}
	\begin{split}
		\rd_{\rm BABAR}\,&=\,0.440\pm0.058\pm0.042  \,,\\ 
		\rd_{\rm BELLE}\,&=\,0.375\pm0.064\pm0.026  \,,\\ 
		\rds_{\rm BABAR}\,&=\,0.332\pm0.024\pm0.018  \,,\\
		\rds_{\rm BELLE}\,&=\,0.293\pm0.038\pm0.015  \,,\\
		\rds_{\rm LHCb}\,&=\,0.336\pm0.027\pm0.030  \,. 
	\end{split}
\end{align}
Here, the first error is statistical and the second one is systematic. Combining these measurements one finds~\cite{BDstaumuCOMBINED}
\begin{align}
	\begin{split}
		\rd_{\rm exp}\,&=\,0.388\pm0.047\,,\\ 
		\rds_{\rm exp}\,&=\,0.321\pm0.021  \,. 
	\end{split}
\end{align}
Comparing the experimental values to the SM prediction\footnote{For these theory predictions we again used the updated results of~\cite{BaBar:2012xj}, which rely on the calculations of Refs.~\cite{Kamenik:2008tj,Fajfer:2012vx} based on the previous results of Refs.~\cite{Korner:1987kd,Korner:1989ve,Korner:1989qb,Heiliger:1989yp,Pham:1992fr}.}
\begin{align}
	\begin{split}
		\rd_{\rm SM}\,&=\,0.297\pm0.017 \,, \\
		\rds_{\rm SM}\,&=\,0.252\pm0.003 \,,
	\end{split}
\end{align}
we see that there is a discrepancy of 1.8$\sigma$ for \rd{} and 3.3$\sigma$ for \rds{}; combining them gives a $3.8 \sigma$ deviation from the SM (compared to $3.4 \sigma$ taking into account the BaBar results only~\cite{Lees:2012xj}).
Models solving the \rd{} puzzle have been discussed extensively in the literature~\cite{Fajfer:2012jt,Crivellin:2012ye,Datta:2012qk,Celis:2012dk,Crivellin:2013wna,Li:2013vlx,Faisel:2013nra,Atoui:2013zza,Sakaki:2013bfa,Dorsner:2013tla,Biancofiore:2014wpa,Alonso:2015sja,Greljo:2015mma,Calibbi:2015kma,Freytsis:2015qca}, including the possibility of charged Higgs particles~\cite{Crivellin:2012ye,Celis:2012dk,Crivellin:2013wna}.

For $\tau\to\mu\nu\nu$, the dominant uncertainty in the SM prediction for the branching ratio comes from the $\tau$ lifetime $\tau_\tau$. Using the PDG~\cite{Olive2014} values for $\tau$ lifetime, $\tau_\tau = (290.3\pm 0.5)\times \unit[10^{-15}]{s}$, and branching ratios 
\begin{align}
	\begin{split}
		B_\mu &\equiv {\rm BR} (\tau\to\mu\overline{\nu}\nu)_\mathrm{exp} =(17.41\pm 0.04)\%\,,\\
		B_e &\equiv {\rm BR} (\tau\to e\overline{\nu}\nu)_\mathrm{exp} =(17.83\pm 0.04)\%\,,
	\end{split}
\end{align}
we can determine the deviations $\Delta_\ell \equiv B_\ell/B_\ell^\mathrm{SM} -1$~\cite{Krawczyk:2004na} from the SM prediction to be
\begin{align}
\Delta_\mu^{\rm PDG} = (0.69\pm 0.29)\%\,, &&
\Delta_e = (0.28\pm 0.28)\%\,.
\label{eq:Deltas}
\end{align}
There is a deviation of about $2.4\sigma$ in the muon data, whereas the electron channel is compatible with the SM prediction. HFAG finds essentially the same value for the tau lifetime, but a slightly lower ${\rm BR} (\tau\to\mu\overline{\nu}\nu)_\mathrm{exp} = (17.39\pm 0.04)\%$~\cite{Amhis:2014hma}, alleviating the deviation to approximately $2\sigma$:
\begin{align}
\Delta_\mu^{\rm HFAG} = (0.59\pm 0.32)\%\,.
\label{eq:DeltasHFAG}
\end{align}
Again, charged Higgses~\cite{Hollik:1992ci,Krawczyk:2004na,Aoki:2009ha} but also neutral $Z^\prime$ bosons~\cite{Crivellin:2015era} affect this decay.

Moving from lepton non-universality to outright lepton flavour violation, we are drawn to the recent CMS excess of $2.4\sigma$ in $h\to\mu\tau$~\cite{Khachatryan:2015kon}:
\begin{align}
	{\rm BR} (h\to\mu\tau) = \left( 0.84_{-0.37}^{+0.39} \right)\% \,.
	\label{h0taumuExp}
\end{align}
Possible explanations naturally require an extended Higgs sector~\cite{Campos:2014zaa,Sierra:2014nqa,Heeck:2014qea,Crivellin:2015mga,Dorsner:2015mja,Omura:2015nja,Crivellin:2015lwa}.

As we see from the previous discussion, all discrepancies outlined above can be solved by additional scalar bosons. In the simplest extension of the SM with additional charged Higgses, a 2HDM, one introduces a second Higgs doublet and obtains four additional physical Higgs particles (in the case of a CP conserving Higgs potential): the neutral CP-even Higgs $H$, a neutral CP-odd Higgs $A$ and the two charged Higgses $H^{\pm}$. 2HDMs have been studied for many years with focus on the type-II models~\cite{Miki:2002nz,WahabElKaffas:2007xd,Deschamps:2009rh}. However, there are also other models without flavour-changing neutral currents at tree-level, i.e.~type-I, type-X (lepton-specific) and the flipped 2HDM (see Ref.~\cite{Branco:2011iw} for an review). More general models with flavour-changing neutral Higgs couplings at tree-level are named type-III models. Here the focus has been on minimal flavour violation~\cite{MFV,Buras:2010mh,Blankenburg:2011ca}, alignment~\cite{Pich:2009sp,Jung:2010ik} or natural flavour conservation~\cite{Glashow:1976nt,Buras:2010mh} but also generic 2HDMs of type III have been studied~\cite{Crivellin:2013wna}. In these models type II is recovered in the absence of flavour-changing neutral Higgs couplings.\footnote{The decoupling limit of the MSSM at tree-level is the 2HDM of type II. However, non-decoupling 1-loop corrections involving the Higgsino mass parameter $\mu$ or non-holomorphic $A$-terms generate ``wrong" Higgs Yukawa couplings giving rise to flavour chaning neutral Higgs couplings (see for example Ref.~\cite{Crivellin:2011jt} for a complete one-loop analysis and Ref.~\cite{Crivellin:2012zz} for the 2-loop SQCD corrections in the MSSM with generic SUSY breaking terms).} While the type-II model cannot explain \rd{} and \rds{} simultaneously~\cite{BaBar:2012xj}, this can be achieved by supplementing the model with additional non-holomorphic couplings~\cite{Crivellin:2012ye}. However, this model (as the normal type-II model) is under pressure from $b\to s\gamma$ data~\cite{Misiak:2015xwa} and LHC searches for $A\to\tau\tau$~\cite{Khachatryan:2014wca}. Furthermore, no sizable effect in the anomalous magnetic moment of the muon can be generated~\cite{Crivellin:2013wna} and also explaining $h\to\mu\tau$ is challenging.

Therefore, we choose to consider in this article the 2HDM of type X (lepton-specific). In this model the heavy Higgs couplings to quarks are suppressed compared to the type-II model and the bounds from $b\to s\gamma$ and LHC searches are much weaker, leaving more space for effects in $(g-2)_\mu$ and tauonic $B$ decays. Furthermore, the sign of the coupling of heavy Higgses to the $\tau$ lepton can be reversed, allowing for constructive interference in $\tau\to\mu\nu\nu$. Also, large effects in $h\to\mu\tau$ compared to the type-II-like model are possible as the Barr-Zee contributions to $\tau\to\mu\gamma$ involving quarks are suppressed.

The article is structured as follows: in the next section we outline our model, i.e.\ the 2HDM of type X with additional non-standard Yukawa couplings. Sec.~3 discusses the relevant observables and collects the necessary formulae. Sec.~4 contains the numerical analysis. Finally we conclude in Sec.~5.

\section{2HDM-X}

We will study a lepton-specific 2HDM (2HDM-X), defined by the Yukawa couplings in the Lagrangian
\begin{align}
	\mathcal{L}_Y = - \overline{Q}_L Y^u \tilde{\Phi}_2 u_R - \overline{Q}_L Y^d \Phi_2 d_R - \overline{L}_L Y^\ell \Phi_1 e_R + \hc \,,
\end{align}
with additional couplings that break the type-X structure
\begin{align}
	\Delta\mathcal{L}_Y = - \overline{Q}_L \xi^u \tilde{\Phi}_1 u_R - \overline{Q}_L \xi^d \Phi_1 d_R - \overline{L}_L \xi^\ell \Phi_2 e_R + \hc
\end{align}
For arbitrary matrices $Y^{u,d,\ell}$ and $\xi^{u,d,\ell}$ this simply parametrizes the 2HDM with the most general Yukawa interactions (type III). We will however assume the coupling structure to be close to type X, i.e.~the $\xi$ matrices to be small perturbations. After electroweak symmetry breaking the following field redefinitions are necessary in order to render the fermion mass matrices diagonal
\begin{equation}
\begin{array}{l}
{d_{L,R}} \to D_{L,R}^\dag {d_{L,R}} \,,\\
{u_{L,R}} \to U_{L,R}^\dag {u_{L,R}} \,,\\
{\ell _{L,R}} \to L_{L,R}^\dag {\ell _{L,R}} \,.
\end{array}
\end{equation}
We define the (non-diagonal) coupling matrices
\begin{align}
	\epsilon^u \equiv U_L^\dagger \xi^u U_R\,, &&
	\epsilon^d \equiv D_L^\dagger \xi^d D_R\,, &&
	\epsilon^\ell \equiv L_L^\dagger \xi^\ell L_R\,,
\end{align}
and express the Yukawa couplings in terms of the physical masses and the couplings $\epsilon$.\footnote{Note that since we eliminate the Yukawa couplings and not the couplings $\xi$ from the Lagrangian, the 2HDM-X is recovered in the limit $\epsilon\to0$.} Note that we are not concerned with the issue of neutrino masses in this article, and hence do not introduce e.g.\ right-handed neutrinos. Therefore, the Higgs interactions with fermions can be written as
\begin{align}
	\begin{split}
		\mathcal{L} \ &\supset \ {{\bar \nu}_i}\Gamma _{{\nu _i}{\ell _j}}^{{H^+ }\;LR}{P_R}{\ell _j}{H^+ }\\
		&\quad+ {{\bar u}_i}\left( {\Gamma _{{u_i}{d_j}}^{{H^+ }\;RL}{P_L} + \Gamma _{{u_i}{d_j}}^{{H^+ }\;LR}{P_R}} \right){d_j}{H^+ }{\mkern 1mu}  \\
		&\quad+\sum\limits_{H_k^0 = {h},{A},{H}}^{} {\sum\limits_{f = u,d,\ell } {\left( {{{\bar f}_i}\, \Gamma _{{f_i}{f_j}}^{H_k^0\;LR} \,{P_R}{f_j}{\mkern 1mu} H_k^0} \right)} } + \hc \,,
	\end{split}
\end{align}
where the couplings are given by
\begin{align}
	\Gamma _{{q_i}{q_j}}^{{h}LR} &\simeq  - \frac{1}{\sqrt2}\left(  \frac{{{m_{{q_i}}}}}{ v}{\delta _{ij}} \cos \alpha- \epsilon _{ij}^q \sin \alpha  \right),\\
	\Gamma _{{q_i}{q_j}}^{{H}LR} &\simeq  - \frac{1}{\sqrt2}\left( \frac{{{m_{{q_i}}}}}{v}{\delta _{ij}} \sin \alpha  + \epsilon _{ij}^q \cos \alpha \right),\\
	\Gamma _{{d_i}{d_j}}^{{A}LR} &\simeq -i\frac{1}{{\sqrt 2 }}\epsilon_{ij}^d \,,\\
	\Gamma _{{u_i}{u_j}}^{{A}LR} &\simeq i\frac{1}{{\sqrt 2 }}\epsilon_{ij}^u\,,\\
	\Gamma _{{u_i}{d_j}}^{{H^+ }LR} &\simeq {V_{ij'}}\epsilon_{j'j}^d\,,\\
	\Gamma _{{u_i}{d_j}}^{{H^+ }RL} &\simeq -\epsilon_{j'i}^{u*}{V_{j'j}}\,,\\
	\Gamma _{{\ell _f}{\ell _i}}^{{h}LR{\kern 1pt} } &\simeq \frac{\sin \alpha \tan \beta}{\sqrt2} \left( {\frac{{{m_{{\ell _i}}}}}{ v}{\delta _{fi}} - {\epsilon_{fi}^\ell }} \right)\,,\\
	\Gamma _{{\ell _f}{\ell _i}}^{{H}LR} &\simeq  - \frac{\cos \alpha \tan \beta}{\sqrt2} \left( {\frac{{{m_{{\ell _i}}}}}{ v}{\delta _{fi}} - {\epsilon_{fi}^\ell }} \right){\mkern 1mu} \,,\\
	\Gamma _{{\ell _f}{\ell _i}}^{{A}LR} &\simeq  - i\frac{\tan \beta }{\sqrt2}\left( {\frac{{{m_{{\ell _i}}}}}{v}\;{\delta _{fi}} - {\epsilon_{fi}^\ell }} \right)\,,\\
	\Gamma _{{\nu _f}{\ell _i}}^{{H^+}LR{\kern 1pt} } &\simeq  \tan \beta \left( {\frac{{{m_{{\ell _i}}}}}{v}{\delta _{fi}} - {\epsilon_{fi}^\ell }} \right)\,,
\end{align}
in the limit of large $\tan\beta$ of interest in this article. $V \equiv U_L^\dagger D_L$ denotes the Cabibbo--Kobayashi--Maskawa (CKM) mixing matrix and $v\simeq \unit[174]{GeV}$ the vacuum expectation value. Note that for $\epsilon^\ell_{33}> m_\tau/v$ the sign of the couplings of $A$, $H$ and $H^+$ to taus is reversed. This will be important later as in this way the sign of the contribution to $\tau\to\mu\nu\nu$ can be flipped. 
In our notation, $h$ is the SM-like Higgs, $H$ and $A$ are the additional CP-even and CP-odd Higges; due to the mainly leptophilic couplings of $H$ and $A$, collider bounds on their masses are quite weak and they can be even lighter than the SM-like scalar boson. In particular, $\sin(\beta-\alpha) = 1$ always corresponds to the SM-like limit, even for $m_H < m_h$ (this differs from standard 2HDM literature).

In the following, we will assume $\epsilon^d=0$ for simplicity, as it is stringently constrained from FCNC processes \cite{Crivellin:2013wna}. In addition, we take $\epsilon^u$ to be of the form
\begin{align}
	\epsilon^u = \matrixx{0 & 0 & 0\\ 0 & 0 & 0\\ 0 & \times & \times } ,
\end{align}
where $\times$ denotes a non-zero entry, since again $\epsilon^u_{13}$ ($\epsilon^u_{23}$) is severely constrained from $b\to d(s)\gamma$ \cite{Crivellin:2011ba}. In the lepton sector we take the same structure as for the $\epsilon^u$:
\begin{align}
	\epsilon^\ell = \matrixx{0 & 0 & 0\\ 0 & 0 & 0\\ 0 & \times & \times } .
	\label{epsilonell}
\end{align}
In this way we avoid lepton flavour violation involving electrons (i.e.~bounds from $\mu\to e\gamma$, $\mu\to eee$ and $\mu\to e$ conversion in nuclei), but still allow for effects in $\Delta a_\mu$, $\tau\to \mu\nu\nu$, and even $h\to\mu\tau$. 

We will not attempt to find a symmetry realisation of the $\epsilon^f$ structures, but take them merely as a convenient minimal set of parameters to explain existing anomalies. Should the above structures prove successful, one might try to find appropriate flavour symmetries to generate them dynamically.

\section{Observables} 

In this section we discuss the relevant processes and summarize the formulae needed for the phenomenological analysis.

\subsection{Tauonic \texorpdfstring{$B$}{B} decays}

In a 2HDM of type II the charged Higgs contribution to \btau{}~\cite{Hou:1992sy}, \rd{} and \rds{} interferes necessarily destructively with the SM and, in addition, \rd{} and \rds{} cannot be explained simultaneously~\cite{BaBar:2012xj}. However, an enhancement is possible in type-III models, see Ref.~\cite{Crivellin:2012ye}. We do not discuss $B\to \tau\nu$ here, which depends on the value of $V_{ub}$. If the inclusive value is taken for $V_{ub}$, it agrees well with the SM prediction, but it is above the SM value if the exclusive determination is used. (Note that the differences between the inclusive and exclusive determination cannot be explained by NP~\cite{Crivellin:2014zpa}.)

\begin{figure}[t]
\centering
\includegraphics[width=0.4\textwidth]{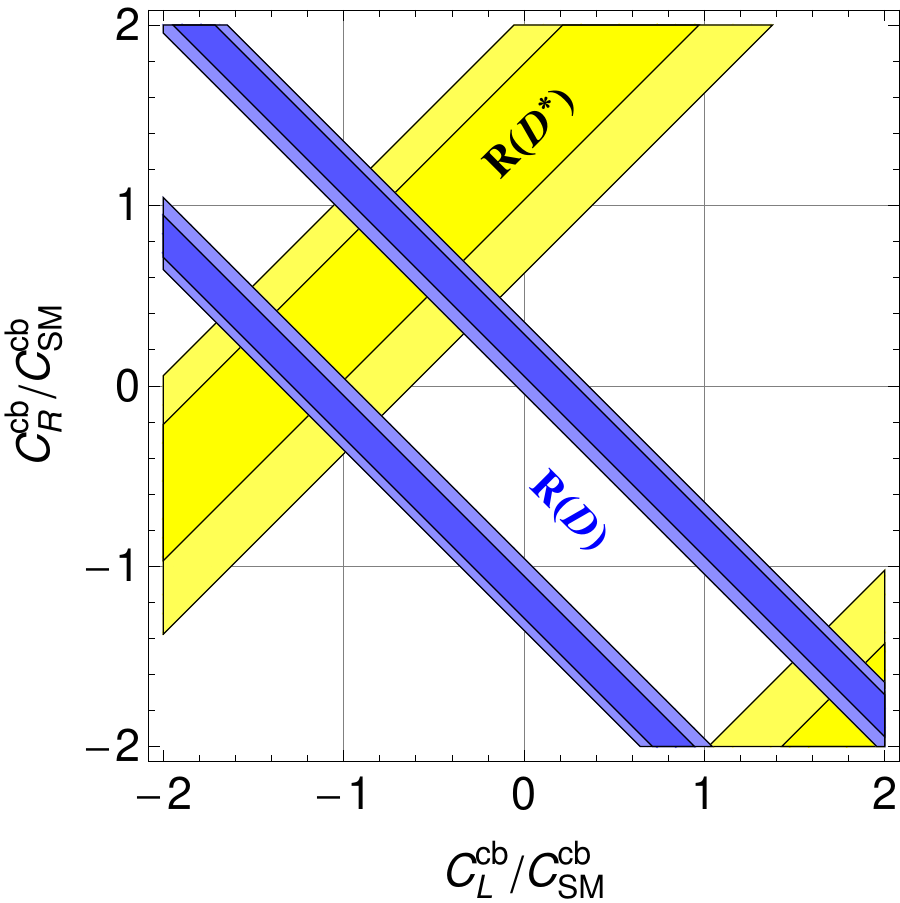}
\caption{Allowed regions in the $C^L_{23}$-$C^R_{23}$ plane from \rd{} (blue) and \rds{} (yellow) for real values of $C^{L,R}_{23}$. The lighter regions correspond to $2\sigma$ experimental uncertainties while the darker regions are correspond to $1\sigma$. 
\label{RDs}}
\end{figure}

The SM and NP contributions relevant for \rd{} and \rds{} are contained in the effective Hamiltonian
\begin{equation}
	{\cal H}_{\rm eff}=  C^{qb}_{\rm SM} O_{{\rm SM}}^{qb} +   C_{R}^{qb} O_{R}^{qb} + C_{L}^{qb} O_{L}^{qb} \,,
\label{Heff}
\end{equation}
with
\begin{align}
	\begin{split}
		O_{{\rm SM}}^{cb}  &= \bar c\gamma _{\mu } P_L b \; \bar\tau \gamma_{\mu } P_L \nu_{\tau}\,, \\ 
		O_{R}^{cb}  &= \bar c P_R b \; \bar\tau P_L \nu_{\tau}\,, \\
		O_{L}^{cb}  &= \bar c P_L b \; \bar\tau P_L \nu_{\tau}\,, 
	\end{split}
\label{Oeff}
\end{align}
assuming massless neutrinos. The SM Wilson coefficient is given by $ C_{{\rm SM}}^{cb} = { 4 G_{F}} \, V^{}_{cb}/{\sqrt{2}}$. 
The NP Wilson coefficients $C_{R}^{cb}$ and $C_{L}^{cb}$ (given at the $B$ meson scale), which parametrize the effect of NP, affect the two ratios in the following way~\cite{Akeroyd:2003zr,Fajfer:2012vx,Sakaki:2012ft}:
\begin{align}
	\frac{\rd}{\rd_{\rm SM}} &=  1 + 1.5 \,\Re\left[\frac{C_{R}^{cb}+C_L^{cb}}{C_{\rm SM}^{cb}}\right]+1.0 \left|\frac{C_{R}^{cb}+C_L^{cb}}{C_{\rm SM}^{cb}}\right|^2  ,\label{RD} \\
	\frac{\rds}{\rds_{\rm SM}} &=  1 + 0.12 \,\Re \left[\frac{C_{R}^{cb}-C_L^{cb}}{C_{\rm SM}^{cb}}\right]+0.05 \left|\frac{C_{R}^{cb}-C_L^{cb}}{C_{\rm SM}^{cb}}\right|^2 .\nonumber
\end{align}
Here, efficiency corrections to \rd{} due to the BABAR detector~\cite{BaBar:2012xj} are important in the case of large contributions from the scalar Wilson coefficients $C_{R,L}^{cb \, ,\tau}$ (i.e.~if one wants to explain ${\cal R}(D)$ with destructive interference with the SM contribution). As shown in Ref.~\cite{Fajfer:2012jt}, these corrections can be effectively taken into account by multiplying the quadratic term in $C_{R,L}^{cb \, ,\tau}$ of \eq{RD} by an approximate factor of 1.5 (not included in \eq{RD}). 

In this model-independent treatment, one can see from Fig.~\ref{RDs} that $C_{R}^{cb}$ alone cannot explain \rd{} and \rds{} simultaneously, whereas $C_{L}^{cb}$ is capable of achieving this, e.g.~with real $C_{L}^{cb} \simeq -1.2 \,|C_{\rm SM}^{cb}|$. In our model, neglecting flavour-changing couplings in the down and in the lepton sector, only the coefficient $C_{L}^{cb}$ is generated in the large $\tan\beta$ limit
\begin{equation}
C_L^{cb} \simeq \frac{{\tan \beta }}{{m_{{H^+ }}^2}}\left( {\epsilon_{32}^{u*} + \epsilon_{22}^{u*}{V_{23}}}  \right)\left( {\frac{{ {m_\tau }}}{v} - \epsilon_{33}^{\ell *}} \right) \,.\end{equation}
For our phenomenological analysis we will neglect $\epsilon^u_{22}$ and add the experimental (statistical and systematic) errors in quadrature, but include the theoretical uncertainty by adding it linearly on top of this.

\subsection{Tau decays \texorpdfstring{$\tau\to\ell\nu\nu$}{tau to l nu nu}}
\label{taumugammaAMM}

\begin{figure}[t]
\centering
\includegraphics[width=0.5\textwidth]{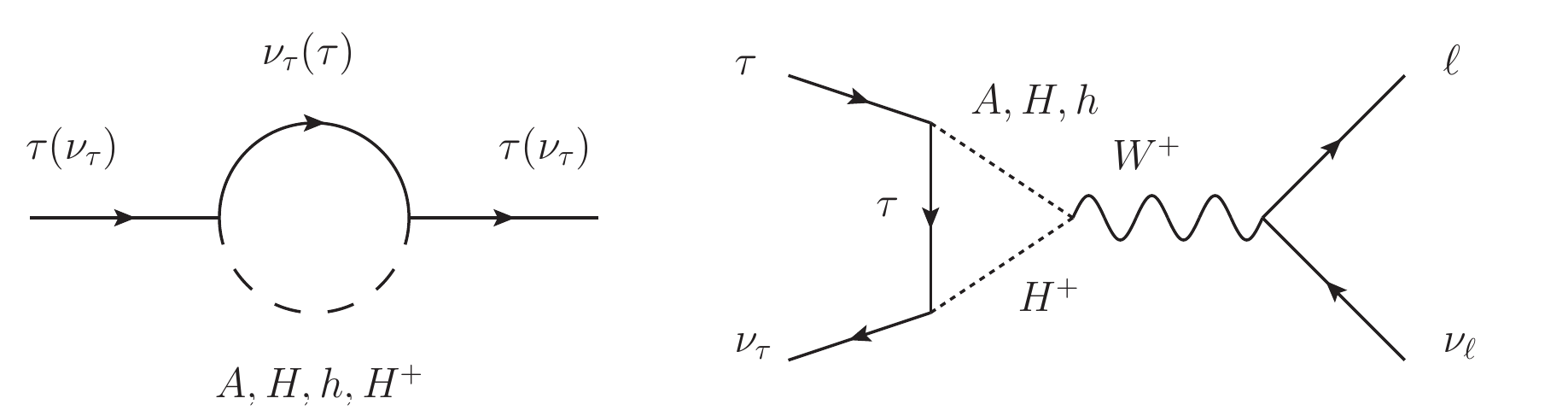}
\caption{Dominant one-loop contributions to $\tau\to\ell \nu\nu$ in the lepton-specific 2HDM (adapted from Ref.~\cite{Krawczyk:2004na}).
\label{2HDM-X_tau_loops}}
\end{figure}

\begin{figure}[t]
\centering
\includegraphics[width=0.4\textwidth]{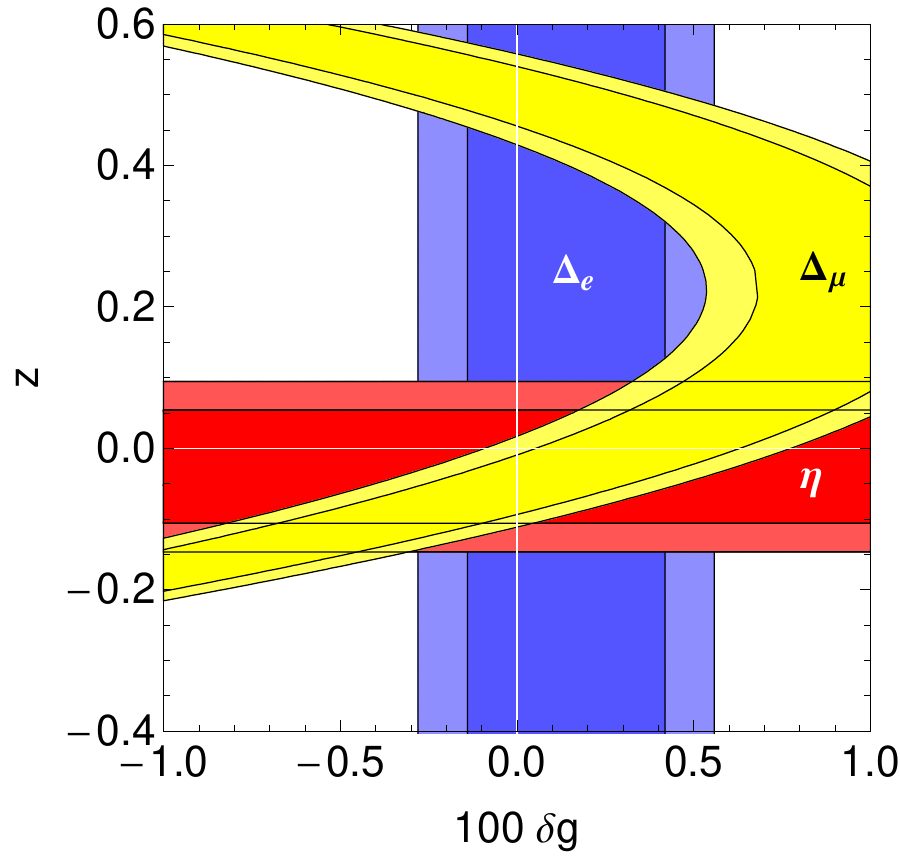}
\caption{Allowed regions for $z$ and $\delta g$ from $\Delta_\mu$ (yellow), $\Delta_e$ (blue), and the Michel parameter $\eta$ (red); see text for definitions. Darker regions are at the $2\sigma$ level, lighter regions at $3\sigma$ one. Here we used for the experimental input the PDG values.
\label{z-delta_g}}
\end{figure}

At tree level, only the charged scalar $H^+$ contributes to $\tau\to\ell\nu\nu$ in the lepton-specific 2HDM. (Note that the contributions in the type-X model are the same as in the type-II model as the lepton sector is identical.) Due to the small electron Yukawa couplings, the contributions to $\Delta_e^\mathrm{tree}$ are highly suppressed compared to $\Delta_e^\mathrm{tree}$, resulting in lepton flavour non-universality~\cite{Hollik:1992ci, Krawczyk:2004na, Aoki:2009ha}. However, one-loop corrections can be important as well (Fig.~\ref{2HDM-X_tau_loops}), and contribute (to a very good approximation) universally to $\Delta_\ell$~\cite{Hollik:1992ci,Krawczyk:2004na} by modifying the $W\tau\nu$ couplings~\cite{Abe:2015oca}
\begin{align}
g_{W\tau\nu}\to g_{W\tau\nu} (1+\delta g)\,.
\end{align}
For $m_\mu\ll m_\tau$ and $\delta g \ll 1$, we find (notation of Eq.~\eqref{eq:Deltas})
\begin{align}
\Delta_e &\simeq 2\delta g\,,\\
\Delta_\mu &\simeq  2\delta g + \frac{z^2}{4} - 2 z \frac{m_\mu}{m_\tau}\,,
\end{align}
with the charged-scalar coupling (assumed to be real)
\begin{align}
z \equiv \frac{v^2}{m_{H^+}^2} \Gamma^{LR\,H^{+}}_{\nu_{\tau}\tau} \Gamma^{LR\,H^{+} \star}_{\nu_{\mu}\mu} \,.
\end{align}
Here we ignored flavour-changing interactions, which would not interfere with the SM and are tightly constrained from flavour-changing neutral current	 processes. In addition, the $H^+$ contribution to $\tau\to\mu\nu\nu$ leads to a non-zero Michel parameter $\eta = -2 z/(4+z^2)$~\cite{Abe:2015oca}, measured to be $\eta = 0.013\pm 0.020$~\cite{Olive2014}. The constraints (using the full expressions for $\Delta_\ell$ including lepton masses~\cite{Abe:2015oca}) are shown in Fig.~\ref{z-delta_g} using PDG values. The SM value is recovered for $\delta g = z = 0$.

In the SM-like limit $\sin(\beta-\alpha)=1$, and ignoring again flavour-violating couplings, we have~\cite{Abe:2015oca}
\begin{align}
\delta g = \frac{\tan^2\beta}{32\pi^2} \left|\frac{m_\tau}{v}-\epsilon^\ell_{33}\right|^2 \left[ F\left(\frac{m_A^2}{m_{H^+}^2}\right)+ F\left(\frac{m_H^2}{m_{H^+}^2}\right)\right] ,
\label{eq:deltag}
\end{align}
with the loop function
\begin{align}
F (x) \equiv \frac12+ \frac{1+x}{4(1-x)} \log x\,.
\end{align}
In particular, $F(x)\leq 0$ (equality for $x=1$), and so $\delta g \leq 0$. In the 2HDM-X, $z$ is positive and $\delta g$ negative, making it hard to reconcile both $\Delta_e$ and $\Delta_\mu$ unless rather high values of $z\simeq 0.5$ are chosen, which are then in disagreement with the Michel parameter $\eta$~\cite{Abe:2015oca}.

In our model, we can however flip the sign of the $\tau$ couplings for $\epsilon^\ell_{33} > m_\tau/v$, which allows for a negative $z$ ($\delta g $ remains negative) as we will see in the phenomenology section. Small values $z\simeq -0.1$ are then sufficient to satisfy the $\Delta_\ell$ constraints as well as $\eta$. Using HFAG values for the $\tau$ decay opens up parameter space at the $2\sigma$ level, see Fig.~\ref{RD_taumununu}.

\subsection{Magnetic moment \texorpdfstring{$a_\mu$}{a(mu)} and \texorpdfstring{$\tau\to \mu\gamma$}{tau to mu gamma}}

The radiative decay $\tau\to \mu \gamma$ is closely related to the anomalous magnetic moment of the muon. Both observables are induced by penguin diagrams with internal neutral or charged Higgs bosons. The results can be encoded in the effective Hamiltonian 
\begin{equation}
\label{HeffLFV}
{\cal{H}}_{\rm eff}=  c^{\ell_{f}\ell_{i}}_{R} \, O^{\ell_{f}\ell_{i}}_{R}
+c^{\ell_{f}\ell_{i}}_{L} \, O^{\ell_{f}\ell_{i}}_{L} \,,
\end{equation}
where $c^{\ell_{f}\ell_{i}}_{R}$ and $c^{\ell_{f}\ell_{i}}_{L}$ are the Wilson coefficients of the magnetic dipole operators
\begin{equation}
O^{\ell_{f}\ell_{i}}_{R(L)}  = m_{\ell_{i}} \bar{\ell}_{f}
\sigma_{\mu \nu} P_{R(L)} \ell_{i} F^{\mu \nu} \, .
\label{HeffLFV2}
\end{equation}
With these conventions, the branching ratio for the radiative lepton
decays $\ell_{i} \to \ell_{f} \gamma$ reads 
\begin{equation}
{\rm BR} (\ell_{i} \to \ell_{f} \gamma )\,=\,\dfrac{m_{\ell_i}^5}{4\pi \, \Gamma_{\ell_i}} \left( |c^{\ell_{f}\ell_{i}}_{R} |^{2}+ |c^{\ell_{f}\ell_{i}}_{L} |^{2} \right ) ,
\label{Brmuegamma}
\end{equation}
and the contribution to the anomalous magnetic moment of the muon is given by
\begin{align}
\delta a_{\mu}=\, -\dfrac{4m^{2}_{\mu}}{e}\, \Re\left[ {c^{\ell_{2}\ell_{2}}_{R} }\right] . 
\end{align}
The one-loop (and dominant two-loop) Wilson coefficients $c_{L,R}$ are given in App.~\ref{sec:wilsons}.

\begin{figure}[t]
\centering
\includegraphics[width=0.4\textwidth]{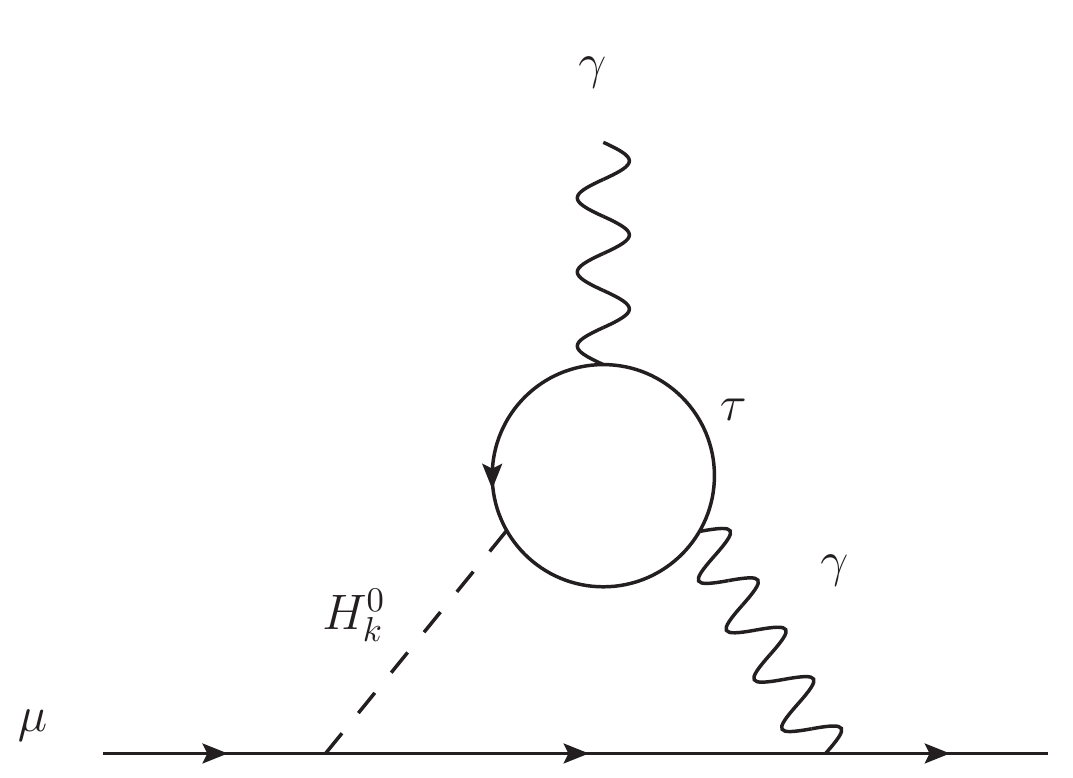}
\caption{Dominating Barr--Zee diagram of a light scalar $H^0_k = A, H$ to $(g-2)_\mu$.
\label{barr-zee}}
\end{figure}

It is well known that two-loop Barr--Zee-type diagrams \cite{Barr:1990vd} can dominate in some regions of parameter space due to enhancement factors $m_f^2/m_\mu^2$ from fermions $f$ in the loop, which can overcome the additional loop suppression $\alpha_\mathrm{EM}/\pi$. For the 2HDM-X, there is an additional $\tan\beta$ enhancement for $A$, $H$, and enhanced two-loop contributions come from the $\tau$ in the loop (Fig.~\ref{barr-zee}). However, there are other important diagrams which could give relevant contributions~\cite{Chang:1993kw,Hisano:2006mj,Jung:2013hka,Abe:2013qla,Ilisie:2015tra}. 
While all diagrams involving $Z$ or $W$ bosons coupling to the external lepton line are highly suppressed, the diagrams with a closed charged Higgs loop or a $W$ loop connected to the external lepton line by a photon could be numerically relevant (replacing the $\tau$ in Fig.~\ref{barr-zee} by a $H^+$ or a $W$). 
Without explicit calculation (the detailed results are given in App.~\ref{sec:wilsons}) one can already deduce the scaling of the relevant diagrams:
\begin{description}
	\item [$\tau$:]$\;$ $\Gamma^{H_k^0LR}_{fi}\frac{m_\tau}{v}\tan^2\beta$ \quad (for $H_k^0=A,H$),
	\item [$H^+$:]$\;$ $\Gamma^{H_k^0LR}_{fi}\lambda_{H^+ H^- H} \tan\beta$ \quad  (for $H_k^0=H$),
	\item [$W$:]$\;$ $\Gamma^{H_k^0LR}_{fi}\cos(\alpha-\beta) \frac{m_W}{v}\tan\beta$ \quad (for $H_k^0=H$).
\end{description}
Here we included only the $A$ and $H$ contributions, as the coupling of $h$ to leptons is suppressed. The couplings of $A$ to $WW$ and $H^+ H^-$ vanish due to CP conservation.

It was pointed out in Ref.~\cite{Ilisie:2015tra} that the Higgs self-coupling $\lambda_{H^+ H^- H}$ contribution can be very important. For this result, Ref.~\cite{Ilisie:2015tra} allowed $\lambda_{H^+ H^- H}$ to vary between $-5$ and $5$. However, $\lambda_{H^+ H^- H}$ is not a fundamental parameter. It depends on the Higgs self-coupling in the scalar potential, but their contribution is suppressed by $1/\tan\beta$ and therefore negligible (see Eq.~\eqref{eq:H+H-H}). The diagram involving a $W$ loop can be important for moderate values of $\tan\beta$~\cite{Chang:1993kw}, but vanishes in the SM-like limit $\sin(\beta-\alpha)=1$. We will see later, $\sin(\beta-\alpha)\neq1$ is required for explaining $h\to\mu\tau$. Nonetheless, working at large $\tan\beta$, as preferred by $\tau$ decays, we checked that also the contribution of the $W$ diagram is numerically small and does not change our conclusion for $(g-2)_\mu$.

\subsection{Flavour changing top decays \texorpdfstring{$t\to H c$}{t to H0 c}}

Since we allow for a non-zero $\epsilon^u_{32}$ coupling, the top quark can decay into $H^0_k+c$ if the scalar $H^0_k =h,H,A$ is sufficiently light. For the branching ratio we find
\begin{equation}
{\rm BR} (t\to H^0_k  c)=\dfrac{m_t}{32\pi\Gamma_t} \left|\Gamma^{H^0_k LR }_{tc}\right|^2 \left(1-\dfrac{m_{H^0_k }^2}{m_t^2}\right) ,
\end{equation}
with the  top decay width $\Gamma_t=(2\pm0.5)\,$GeV. 

\subsection{Leptonic Higgs decays}
\label{sec:higgsdecays}

The decay $h\to \tau\tau$ has been observed by CMS~\cite{Chatrchyan:2014nva} and ATLAS~\cite{Aad:2015vsa} with $\mu$-parameters (relative strength compared to the SM prediction) $0.78\pm 0.27$ and $1.43^{+0.43}_{-0.37}$, respectively. A naive combination (also averaging the ATLAS result to $1.43\pm 0.40$) gives
\begin{align}
\mu_{\tau\tau} = 0.98\pm 0.22 \,.
\end{align}
In our model, the decay rate for $h\to \tau\tau$ relative to the SM prediction takes the form
\begin{align}
\mu_{\tau\tau} \simeq\frac{{\rm BR}\left(h\to\tau\tau\right)}{{\rm BR}\left(h\to\tau\tau\right)_\text{SM}} = \sin^2\alpha \tan^2\beta \left| 1- \frac{\epsilon^\ell_{33}}{m_\tau/v}\right|^2 ,
\end{align}
in the large $\tan\beta$ limit, assuming a SM-like production rate (in the SM-like limit $\sin(\beta-\alpha)\to 1$ we get back $\mu_{\tau\tau}\to 1$).
The entry $\epsilon^\ell_{32}$ allows for the flavour-violating decay $h\to\mu\tau$ if $\cos(\alpha-\beta)\neq 0$:
\begin{align}
{\rm BR}&\left(h\to\mu\tau\right) \simeq \dfrac{m_{h}}{8\pi \Gamma_{h}} \left| {\Gamma^{h\,LR}_{\tau\mu}} \right|^2 ,
\end{align}
where $\Gamma_{h}\simeq 4.1\,$MeV is the decay width in the SM for the $125\,$GeV Brout--Englert--Higgs boson.

\subsection{Other Higgs decays}

\begin{figure}[t]
\centering
\includegraphics[width=0.4\textwidth]{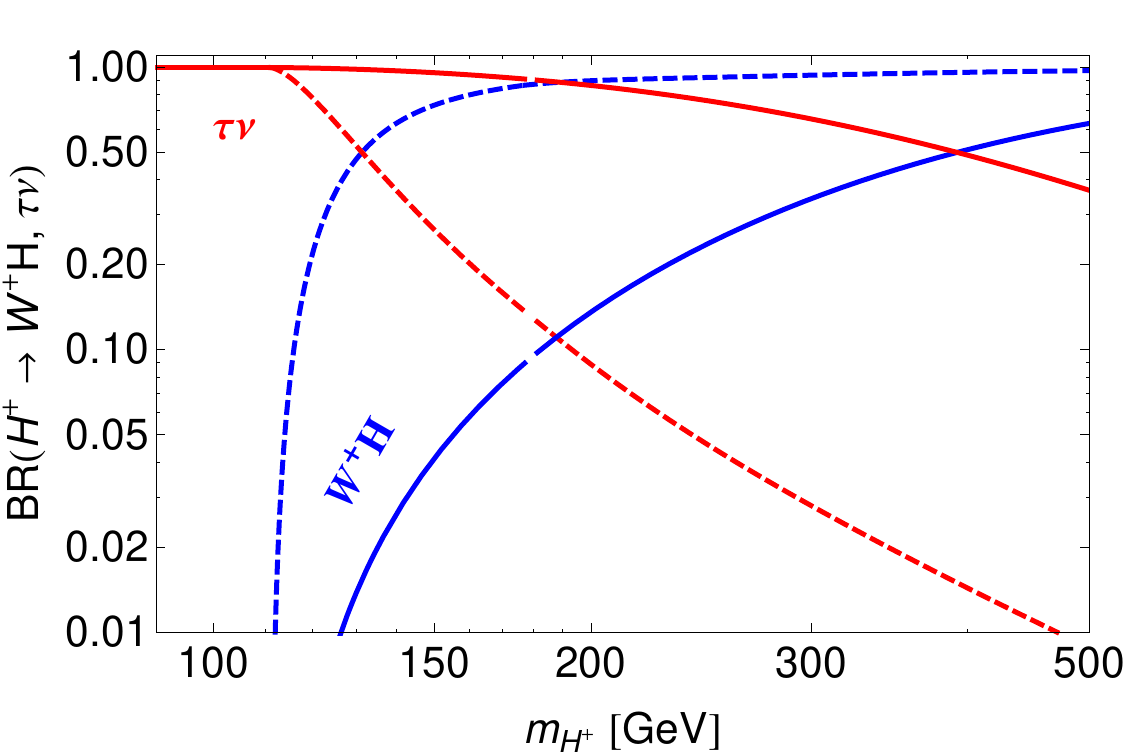}
\caption{Branching ratios $H^+\to W^+ H$ (blue) and $H^+\to\tau\nu$ (red) for $m_H = \unit[30]{GeV}$. Solid (dashed) lines are for $\tan\beta = 80$ $(10)$.
\label{charged_higgs_decay}}
\end{figure}

With a light $H$, new decay modes open up, such as $h\to H H$, $A\to Z H$, and $H^+\to W^+ H$~\cite{Gunion:1989we}, followed by $H \to \tau\tau$ in the large $\tan\beta$ limit. (In complete analogy to the case where $A$ is light, see e.g.~\cite{Cao:2009as}.) The decay rates are given by
\begin{align}
\Gamma (H^+\to W^+ H) &\simeq \frac{\alpha_\mathrm{EM}}{48 s_W^2} \frac{m_{H^+}^3}{m_W^2} \left(1- \frac{m_W^2}{m_{H^+}^2}\right)^3 ,\\
\Gamma (A\to Z H) &\simeq \frac{\alpha_\mathrm{EM}}{48 s_W^2 c_W^2} \frac{m_{A}^3}{m_Z^2} \left(1- \frac{m_Z^2}{m_{A}^2}\right)^3 ,
\end{align}
in the SM-like limit $\sin (\beta-\alpha)\simeq 1$ and for $m_H \ll m_W$. ($h\to H H $ depends on additional parameters and will not be discussed here.)
If $\tan\beta$ is not too large, these channels can contribute significantly to the total decay rates of $H^+$ and $A$ (see Fig.~\ref{charged_higgs_decay}), which weakens the direct search bounds on these particles (these bounds often assume 100\% decays into tau leptons for the 2HDM-X).

\subsection{Z boson decays}

In the large $\tan\beta$ limit one expects a sizable modification of $Z\to \tau\tau$ at the loop level.
Following Ref.~\cite{Abe:2015oca} we define the ratio of decays
\begin{align}
R_{\tau/e} \equiv \frac{\Gamma (Z\to\tau\tau)}{\Gamma (Z\to ee)},
\end{align}
with the experimental value $R^\text{exp}_{\tau/e} = 1.0019\pm 0.0032$~\cite{Olive2014}. The deviation from the SM due to 2HDM vertex corrections is given by
\begin{align}
\begin{split}
\Delta R_{\tau/e} &\equiv R_{\tau/e}-R_{\tau/e}^\text{SM}\\
&\simeq \frac{g_Z^2 m_Z/6\pi}{\Gamma (Z\to \ell\ell)_\text{SM}} \left[ v_\tau \Re \Delta v_\tau^\text{loop} + a_\tau \Re \Delta a_\tau^\text{loop} \right] ,
\end{split}
\end{align}
with $v_\tau = \frac12 T_3 - s_W^2 Q_\tau = s_W^2 -1/4$ and $a_\tau = \frac12 T_3 = -1/4$ being the tree level $Z\tau\tau$ couplings and
\begin{align}
\begin{split}
\Delta v_\tau^\text{loop} &\simeq \frac{ \tan^2\beta}{32\pi^2} \left(\frac{m_\tau}{ v} - \epsilon^\ell_{33}\right)^2\left[ v_\tau (F_1 (m_H) + F_1 (m_A))\right.\\
&\left.\quad -2 a_\tau F_1 (m_{H^+}) + (v_\tau+a_\tau) F_2 (m_{H^+},m_{H^+})\right] ,\\
\Delta a_\tau^\text{loop} &\simeq -\frac{ \tan^2\beta}{32\pi^2} \left(\frac{m_\tau}{ v}- \epsilon^\ell_{33}\right)^2 \left[ a_\tau ( F_1 (m_H) + F_1 (m_A))\right.\\
&\left.\quad -2 a_\tau F_1 (m_{H^+}) + (v_\tau+a_\tau) F_2 (m_{H^+},m_{H^+})\right.\\ 
&\left.\quad -4 a_\tau F_2 (m_{H},m_{A})\right] ,
\end{split}
\end{align}
in the SM-like limit $\sin (\beta -\alpha) = 1$. The loop functions $F_j$ can be found in Ref.~\cite{Abe:2015oca}.
For the region of interest the limits from $Z$ decays are of similar order as those from $\tau\to \ell\nu\nu$.

For $\epsilon_{32}^\ell \neq 0$, one also induces $Z\to\mu\tau$ via a similar one-loop diagram~\cite{Goto:2015iha}.
For the mass ranges of interest in this work the branching ratio is approximately
\begin{align}
{\rm BR}(Z\to\mu\tau) \simeq 5\times 10^{-5} \left(\frac{\epsilon_{32}^\ell}{10^{-2}}\right)^2 \left(\frac{\tan\beta}{100}\right)^4 .
\end{align}
The best upper limit of $1.2\times 10^{-5}$ at $95\%$~C.L.~\cite{Olive2014} still comes from LEP, but LHC searches should be able to improve this by a factor of few with current data~\cite{Davidson:2012wn} and even more with the upcoming run.

\section{Phenomenological Analysis}

Using the formulae collected in the last section, we now study the phenomenology of our 2HDM and show that it can indeed resolve the anomalies outlined in the introduction.

\subsection{\texorpdfstring{$\rd$}{R(D)} and \texorpdfstring{$\rds$}{R(D*)}}

\begin{figure*}[t]
\centering
\includegraphics[height=0.4\textwidth]{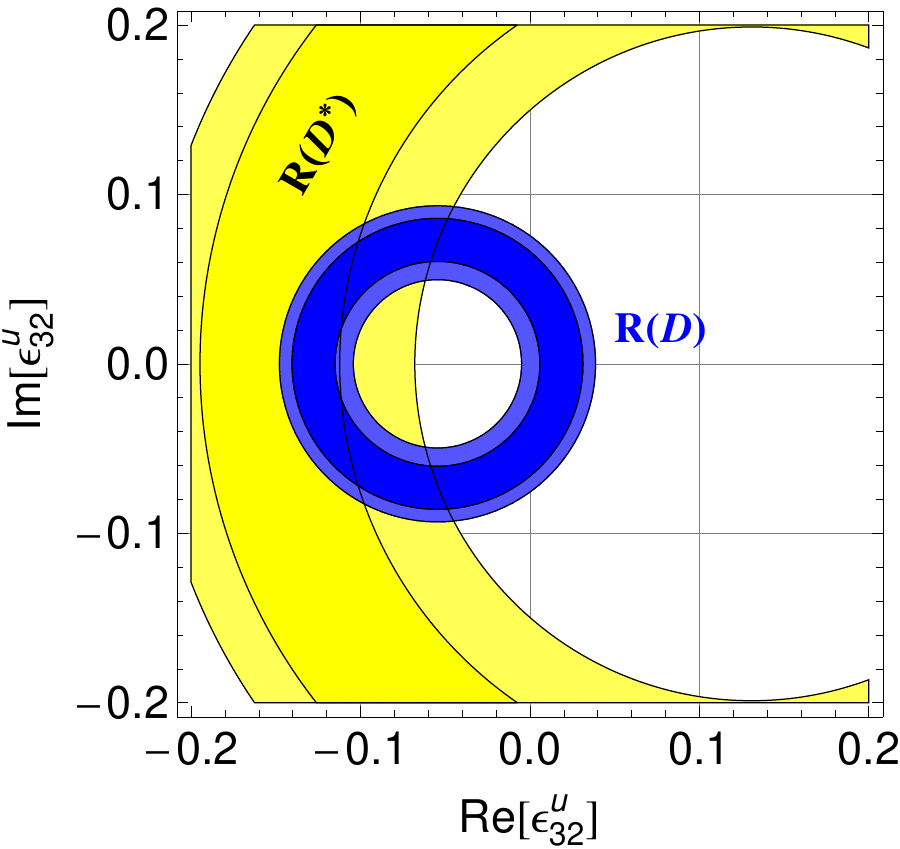}\hspace{2ex}
\includegraphics[height=0.4\textwidth]{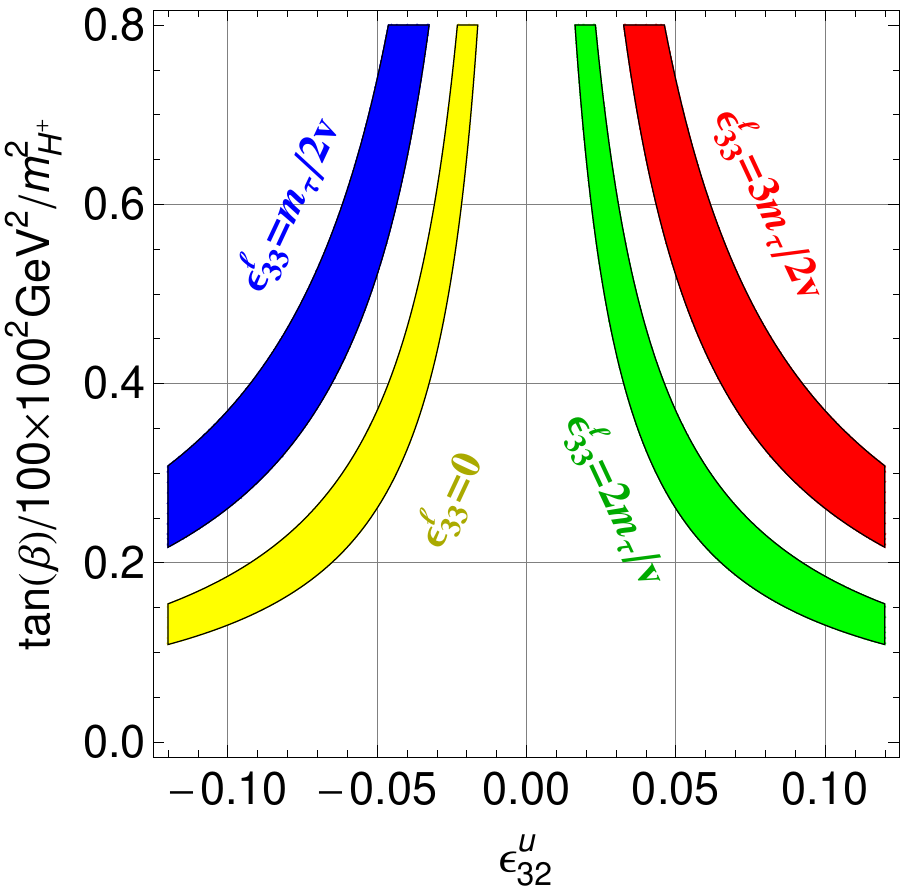}
\caption{Left: Allowed regions in the $\epsilon^u_{32}$-plane from \rd{} (blue) and \rds{} (yellow) for $\tan\beta=50$, $m_{H^+}=200$~GeV, and $\epsilon^\ell_{33} = 0$. The scaling of the allowed region for $\epsilon^u_{32}$ with $\tan\beta$ and $m_{H^+}$ is the same as for $\epsilon^u_{31}$. $\epsilon^u$ is given at the matching scale $m_{H^+}$. 
Right: Allowed regions in the $\epsilon	^u_{32}$--$\dfrac{\tan\beta}{100}\dfrac{(100\,{\rm GeV})^2}{m_{H^+}^2}$ plane for $\epsilon^\ell_{33}=0$ (yellow), $\dfrac{m_\tau}{2v}$ (blue), $\dfrac{3m_\tau}{2v}$ (red) and $\dfrac{2m_\tau}{v}$ (green). 
\label{xiu32}}
\end{figure*}

Let us first consider $\rd$ and $\rds$. From the left plot of Fig.~\ref{xiu32} we see that $\rd$ and $\rds$ can be explained simultaneously for negative  values of $\epsilon^u_{32}$ with small or vanishing imaginary part. The right plot in Fig.~\ref{xiu32} shows the dependence of $\epsilon^u_{32}$ on $\dfrac{\tan\beta}{100}\dfrac{(100\,{\rm GeV})^2}{m_{H^+}^2}$ requiring that $\rd$ and $\rds$ are explained. Note that sizable values of $\epsilon^u_{32}$ are required, i.e.\ of the order of $10^{-1}$. This is important for $t\to H c$ to be considered later.

\subsection{\texorpdfstring{$\tau\to \ell\nu\nu$}{tau to l nu nu}}

\begin{figure}[t]
\centering
\includegraphics[width=0.45\textwidth]{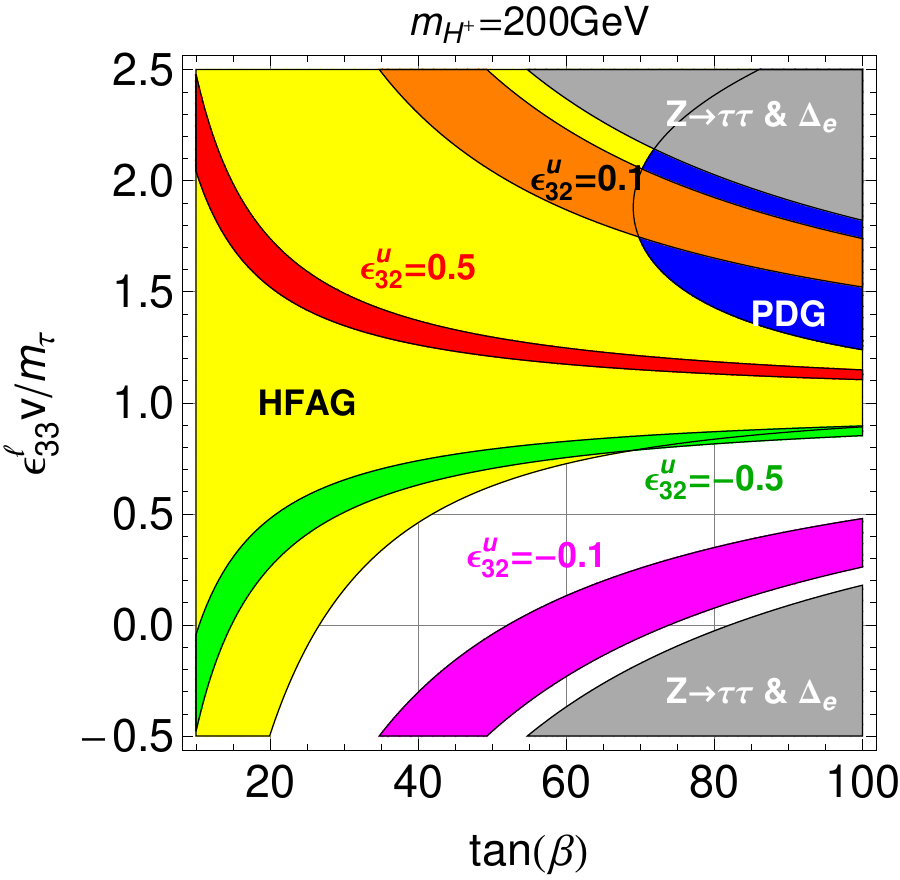}
\caption{Allowed regions in the $\tan\beta$--$v/m_\tau\epsilon^\ell_{33}$ plane from \rdrds{} and $\tau\to\mu\nu\nu$ at the $2\,\sigma$ level. The yellow region is allowed by $\tau\to\mu\nu\nu$ using the HFAG result for $m_H=30\,$GeV and $m_A=200\,$GeV, while the (darker) blue one is the allowed region using the PDG result. The red, orange, green and magenta bands correspond to the allowed regions by \rdrds{} for different values of $\epsilon^u_{32}$. The gray region is excluded by $Z\to \tau\tau$ and $\tau\to e \nu\nu$. For $m_H\approx m_A$ the allowed regions from $\tau\to\mu\nu\nu$ would be slightly larger.
\label{RD_taumununu}}
\end{figure}

The tree-level charged Higgs contribution interferes destructively with the SM for $\epsilon_{33}^\ell=0$. However, for $\epsilon_{33}^\ell> m_\tau/v$ the interference is constructive, allowing for an explanation of the PDG data, which is in more than a $2\sigma$ tension with the SM. The 1-loop contributions interfere again destructively (independently of $\epsilon_{33}$) and are important for non-degenerate $A$ and $H$ masses. Nonetheless, even if $m_H=30\,$GeV and $m_A=200\,$GeV, the values $m_{H^+}=200\,$ GeV, $\epsilon_{33}^\ell=2 m_\tau/v$ and $\tan\beta>60$ are consistent with data (see Fig.~\ref{RD_taumununu}). Furthermore, as one can also see from Fig.~\ref{RD_taumununu}, for $\epsilon^u_{32}\approx 0.1$ also $\rd$ and $\rds$ can be brought into agreement with the measurements. Therefore, the possibility to flip the sign of the $H,A$ coupling to taus allows us to have smaller values $\tan\beta$ than in the 2HDM-X.

\subsection{Anomalous magnetic moment of the muon}

\begin{figure*}
\centering
\includegraphics[width=0.45\textwidth]{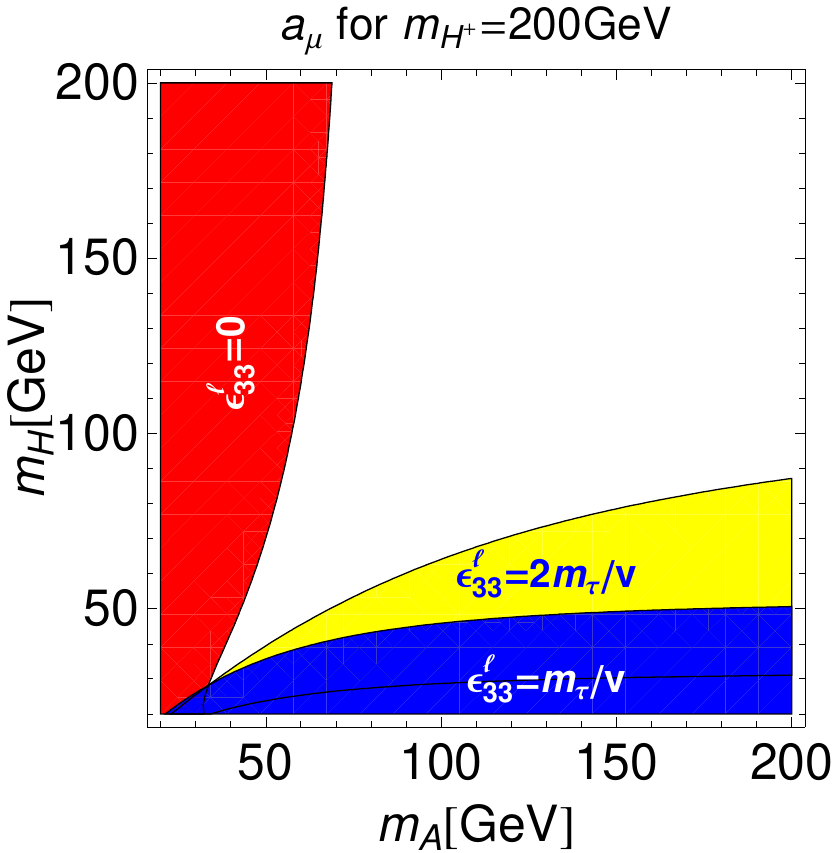}
\includegraphics[width=0.45\textwidth]{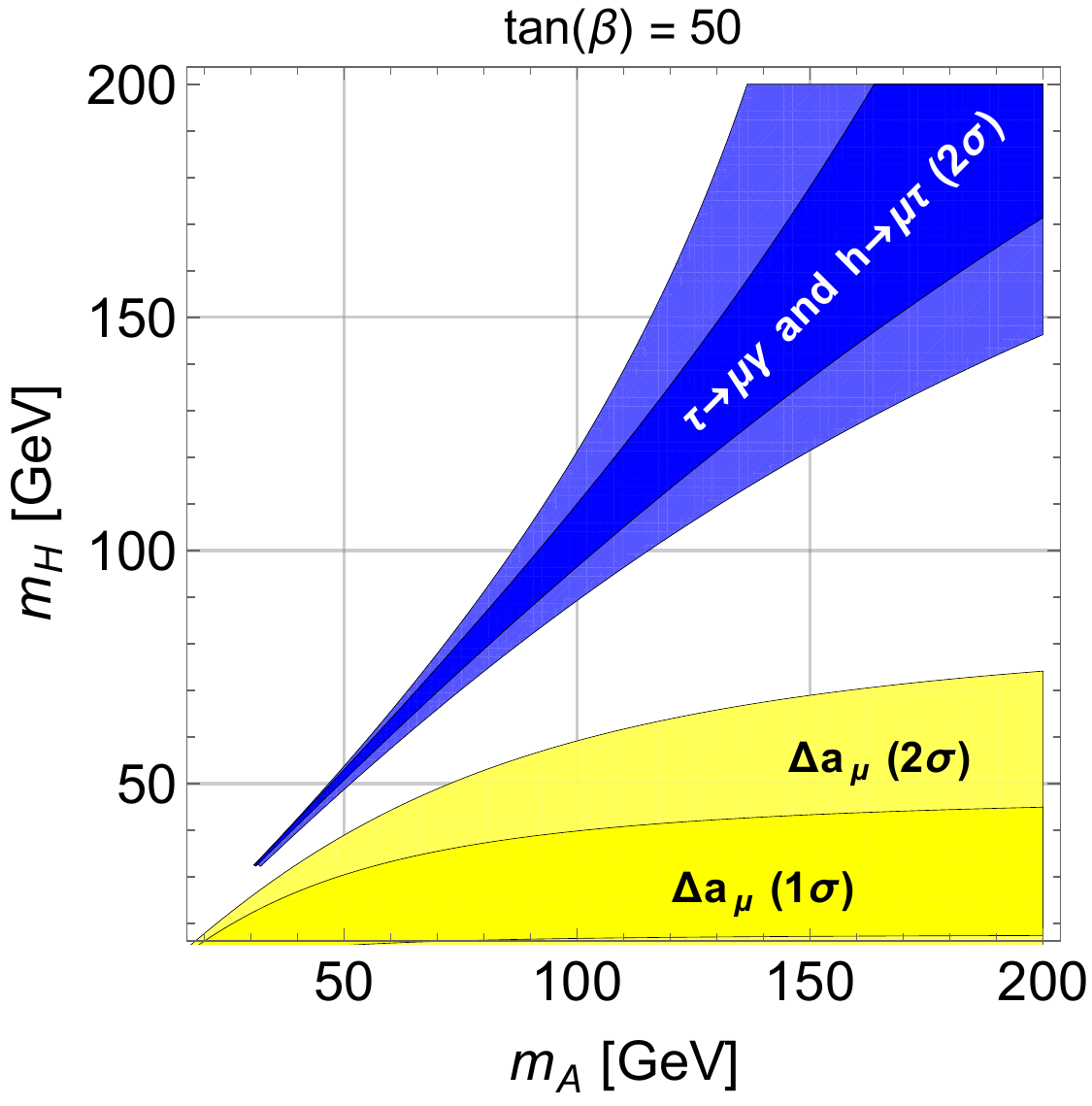}
\caption{Left: Allowed regions in the $m_A$--$m_H$ plane from the anomalous magnetic moment of the muon at the $2\,\sigma$ level for $\tan\beta=70$, $m_{H^+}=200\,$GeV, $\cos(\alpha-\beta)=0$ and different values of $\epsilon^\ell_{33}$. 
Right: Allowed regions in the $m_A$-$m_H$ plane from $(g-2)_\mu$, $\tau\to\mu\gamma$ and $h\to\mu\tau$ for $\tan(\beta)=30$ and $\epsilon^\ell_{33} = 2m_\tau/v$. For $h\to\tau\mu$ blue corresponds to $\cos(\alpha-\beta)=0.1$ and light blue to $\cos(\alpha-\beta)=0.2$. The allowed region for $\Delta a_\mu$ is maximal in the sense that we allowed for the three possibilities $\epsilon^\ell_{32}\neq0$, $\epsilon^\ell_{32}=\epsilon^\ell_{23}\neq0$ and $\epsilon^\ell_{32}=-\epsilon^\ell_{23}\neq0$ as the latter ones can give $m_\tau/m_\mu$ enhanced one-loop contributions. However, the effects turns out to be small, as $\epsilon^\ell_{32,32}$ is stringently constrained from $\tau\to\mu\gamma$. In addition, we checked that the effect of $\lambda_H$ is very small. 
\label{AMM_mA-MH}}
\end{figure*}

In the anomalous magnetic moment of the muon, the one-loop and the two-loop Barr--Zee contribution have opposite sign for $\epsilon^\ell_{33}=0$ (neglecting flavour violating couplings). However, for $\epsilon^\ell_{33}> m_\tau/v$ the interference is constructive, allowing for an explanation with smaller values of $\tan\beta$ and/or heavier Higgses. Note that for $\epsilon^\ell_{33}> m_\tau/v$ the $H$ contribution has the same sign as the SM contribution while the $A$ one has opposite sign, so in our scenario it is a light $H$ that can solve the $\Delta a_\mu$ anomaly, as opposed to a light $A$ in the standard 2HDM-X. We show explicitly in the left plot of Fig.~\ref{AMM_mA-MH} that $m_{H}$ must be smaller than $m_A$ for $\epsilon^\ell_{33}> m_\tau/v$. As seen above, $\epsilon^\ell_{33}> m_\tau/v$ is preferred by $\tau\to\mu\nu\nu$.

\subsection{\texorpdfstring{$h\to\mu\tau$}{h to mu tau} and \texorpdfstring{$\tau\to \mu\gamma$}{tau to mu gamma}}

Until now, we worked in the large $\tan\beta$ limit with $\alpha=0$. However, the decay $h\to\mu\tau$ can only appear for non-zero values of $\cos(\alpha-\beta)$. In this case additional Barr--Zee diagrams with gauge bosons or top quarks can contribute to $\tau\to\mu\gamma$ (see App.~\ref{sec:wilsons}). Therefore, the analysis is more involved than the one for the anomalous magnetic moment of the muon. However, as we have shown in the case of $(g-2)_\mu$ (where the contributions are directly related to $\tau\to\mu\gamma$), $\cos(\alpha-\beta)\neq0$ has actually only a small effect on the result.

To explain the CMS excess in $h\to\mu\tau$ (Eq.~\eqref{h0taumuExp}), one needs a coupling strength of approximately
\begin{align}
\sin\alpha \tan\beta |\epsilon^\ell_{32}|\simeq 3.7\times 10^{-3}\,.
\end{align}
Non-zero values of $\epsilon^\ell_{32}$ then give rise to $\tau \to \mu \gamma$. The experimental upper limit for $\tau \to \mu \gamma$ is given by~\cite{Aubert:2009ag,Hayasaka:2007vc} 
\begin{equation}
	{\rm BR}\left( \tau \to \mu \gamma \right)  \leq 4.4 \times 10^{-8}\,,
\end{equation}
at $90\%$~C.L.. It is interesting to see if one can explain $h\to \mu\tau$ and $(g-2)_\mu$ simultaneously without violating bounds from $\tau\to\mu\gamma$. As the loop contributions to $\tau\to\mu\gamma$ are governed by the same Wilson coefficients as $(g-2)_\mu$ (see Sec.~\ref{taumugammaAMM}), this turns out to be challenging. In the right plot of Fig.~\ref{AMM_mA-MH} we show the allowed regions in the $m_A$--$m_H$ plane for $\tau\to\mu\gamma$, $h\to \mu\tau$ and $(g-2)_\mu$. As one can see, there is no overlap among all regions. There is an $m_\tau/m_\mu$ enhanced contribution to $(g-2)_\mu$ in the case $\epsilon_{23}^\ell\neq0$ and $\epsilon_{32}^\ell\neq0$. Even though we restricted ourselves in \eq{epsilonell} to vanishing values of $\epsilon_{23}^\ell\neq0$, we checked that also for $\epsilon_{32}^\ell=\epsilon_{23}^\ell$ the effect in $(g-2)_\mu$ is small, taking into account the upper limit on $\epsilon_{32}^\ell=\epsilon_{23}^\ell$ from $\tau\to\mu\gamma$ while aiming at an explanation of $h\to\mu\tau$.

In principle, one might increase the coupling strength $\Gamma^{H,A,H^+}_{\mu\mu}$ with the help of $\epsilon_{22}^\ell$. This would soften the tight relationship 
\begin{align}
c_R^{\mu\tau}/c_R^{\mu\mu} \simeq \Gamma^{H^0_k}_{\mu\tau}/\Gamma^{H^0_k}_{\mu\mu} \simeq \epsilon^\ell_{32} v/m_\mu
\end{align}
originating from the dominant Barr--Zee diagram with a $\tau$ loop which causes the incompatibility of $a_\mu$ and $\tau\to\mu\gamma$ (Fig.~\ref{barr-zee}). However, a large shift in $\Gamma^{H,A,H^+}_{\mu\mu}$ from $\epsilon^\ell_{22}\ll m_\mu/v$ would mean fine tuning and also strongly affect $\tau\to\mu\nu\nu$. Therefore, we conclude that explaining $a_\mu$ and $h\to\mu\tau$ simultaneously is not impossible, but rather difficult and would involve fine tuning.

\subsection{\texorpdfstring{$t\to H c$}{t to H c}}

\begin{figure}[b]
\centering
\includegraphics[width=0.45\textwidth]{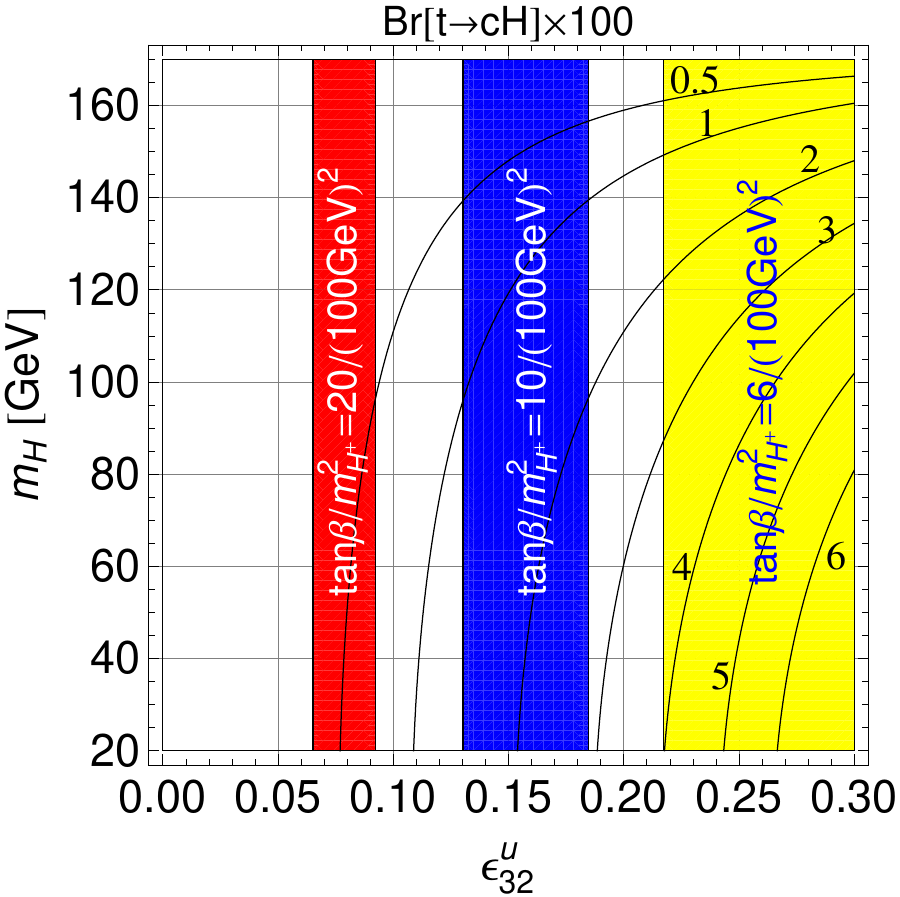}
\caption{The contour lines denote ${\rm BR}(t\to H c)\times 100$ as a function of $\epsilon^u_{32}$ and $m_H$. The coloured regions are allowed by \rd{} and \rds{} for different values of $\tan\beta/m_{H^+}^2$.
\label{topHc}}
\end{figure}

For light values of $m_H$, as preferred by the anomalous magnetic moment of the muon, and non-zero values of $\epsilon^u_{32}$ as required by an explanation of $\rdrds$, the flavour changing top decay $t\to H c$ can have sizable branching ratios. In fact, as shown in Fig.~\ref{topHc} the branching ratio can be easily of the order of $10^{-2}$.

%\clearpage
\section{Conclusions}

In this article we addressed the measured anomalies in \rdrds{} ($3.8\sigma$), $a_\mu$ ($\sim3\sigma$), $\tau\to\mu\nu\nu$ ($2.4\sigma$), and $h\to\mu\tau$ ($2.4\sigma$) within a simple two-Higgs-doublet model. The Yukawa structure of our model is close to the lepton-specific 2HDM (type X), but with some additional Yukawa couplings involving third-generation fermions that give rise to the $b$--$c$ (necessary for \rdrds{}) and $\mu$--$\tau$ transitions (relevant for $h\to\mu\tau$) as well as corrections to $\tau\tau$ couplings (important for $\tau\to\mu\nu\nu$).

Let us summarize the logic of the article.
\begin{itemize}
	\item $\tau\to\mu\nu\nu$ prefers $\epsilon^\ell_{33}\geq m_\tau/v$. If one wants to account for the PDG average, also $\tan\beta>50$ is desirable.
	\item $a_\mu$ favours small values of $m_H$ for $\epsilon^\ell_{33}\geq m_\tau/v$ as in this case the Barr--Zee contribution with a $\tau$ loop  has the correct sign and the diagram involving a Higgs self-coupling can be relevant.
	\item \rd{} and \rds{} point towards quite large negative values of $\epsilon^u_{32}$ (of order $-0.1$).
	\item In case of a light $H$ (as preferred by $a_\mu$), sizable decay rates for $t\to Hc$ are possible if \rd{} and \rds{} are explained via $\epsilon^u_{32}$. This decay could be observed at the LHC in the process $pp\to\bar tt$, $t\to Hc$, $H\to\tau\tau$.
	\item $h\to\mu\tau$ can be explained using $\epsilon^\ell_{32}$. In this case $\alpha\neq0$ and constraints from $\tau\to\mu\gamma$ arise. As the Barr--Zee contributions in $\tau\to\mu\gamma$ are directly correlated to the ones in $a_\mu$, a simultaneous explanation is difficult.
	\item If one attempts to explain $h\to\mu\tau$ (disregarding $a_\mu$), the exotic process $pp\to\bar tt$, $t\to Hc$, $H\to\mu\tau$ can occur at the LHC.
	\end{itemize}

Therefore, the future prospects are very promising: the decay $h\to\mu\tau$ implies rates for $\tau\to \mu\gamma$ in reach of future experiments. More data on tau leptons is necessary to test our model, in particular $\tau\to\ell\nu\nu$, $\tau\to\mu\gamma$, and $h\to\tau\tau$. Furthermore, the light $H$ and the flavour-changing couplings required for \rdrds{} lead to the decay $t\to H c$, followed by $H\to \tau\tau$ (or even $H\to \mu\tau$), which can be searched for at the LHC. While we did not attempt to find a symmetry realisation of the pattern assumed for the $\epsilon^f_{ij}$ structures, it would be very interesting to find appropriate flavour symmetries to generate them dynamically, as the model works very well phenomenologically.
An additional venue of interest would be the inclusion of dark matter in order to explain the galactic centre gamma-ray excess~\cite{Hektor:2015zba}.

%%%%%%%%%%%%%%%%%%%%%%%%%%%%%%
\vspace{2mm}
\acknowledgments{A.~Crivellin is supported by a Marie Curie Intra-European Fellowship of the European Community's 7th Framework Programme under contract number PIEF-GA-2012-326948. The work of J.~Heeck is funded in part by IISN and by Belgian Science Policy (IAP VII/37). P.~Stoffer gratefully acknowledges financial support by the DFG (CRC 16, ``Subnuclear Structure of Matter'').}

%\clearpage
\appendix
\section{Wilson coefficients for \texorpdfstring{$a_\mu$}{a(mu)} and \texorpdfstring{$\ell_i\to \ell_f\gamma$}{li to lf gamma}}
\label{sec:wilsons}

The effective Hamiltonian relevant for $\ell_i \to \ell_f \gamma$ and $a_\mu$ is given in Eq.~\eqref{HeffLFV} with operators from Eq.~\eqref{HeffLFV2}. The Hermiticity of the Hamiltonian allows us to deduce $c_L$ from $c_R$ in complete generality via
\begin{equation}
	c^{\ell_{f}\ell_{i}}_{L}= \frac{m_{\ell_f}}{m_{\ell_i}} c^{\ell_{i}\ell_{f}*}_{R}\,,
	\label{cLcR}
\end{equation}
so we will only show $c_R$ in the following. The final $c_{L,R}$ of course requires a sum over all the individual $c_{L,R}$ we present here.

At one loop the neutral Higgs ($H_{k}^{0}=H,h,A$) penguin contribution to $c^{\ell_{f}\ell_{i}}_{R}$ is given by
\begin{align}
\begin{split}
{c^{\ell_{f}\ell_{i}}_{R\,H^0}}  &=   \sum\limits_{k,j = 1}^3  
 \dfrac{-e}{192 \pi^2  m^2_{H^{0}_{k}} }\, \Bigg[  \Gamma^{H_{k}^{0}\,LR}_{\ell_{f}    \ell_{j}}\Gamma^{H_{k}^{0}\,LR
    \star}_{\ell_{i} \ell_{j}}  \\		&\quad+\dfrac{m_{\ell_{f}}}{m_{\ell_{i}}}
  \Gamma^{H_{k}^{0}\,LR \star}_{\ell_{j} \ell_{f}}  \Gamma^{H_{k}^{0}\,LR}_{\ell_{j} \ell_{i}}   \\
  &\quad -\dfrac{m_{\ell_{j}}}{m_{\ell_{i}}} \Gamma^{H_{k}^{0}\,LR}_{\ell_{f}    \ell_{j}}   \Gamma^{H_{k}^{0}\,LR}_{\ell_{j} \ell_{i}}      \left(
  9+ 6 \ln\left(\dfrac{   m^{2}_{\ell_{j}}  }{m^2_{H^{0}_{k}}}\right)  \right)      \Bigg]  ,
\end{split}
\label{CiH0klitolfgamma} 
\end{align}
and the charged Higgs penguin contribution takes the form
\begin{align}
  {c^{\ell_{f}\ell_{i}}_{R\,H^{+}}} &=   \dfrac{e}{384 \pi^2   m^2_{H^{+}} } \frac{m_{\ell_f}}{m_{\ell_i}}  \sum\limits_{j = 1}^3  \Gamma^{H^{+}\,LR}_{\nu_{j}\ell_{i}}  \Gamma^{H^{+}\,LR
    \star}_{\nu_{j}\ell_{f}}  \,.
\end{align}

\begin{figure*}[t]
	\centering
	\subfloat[]{
		\centering
		\includegraphics[width=4cm]{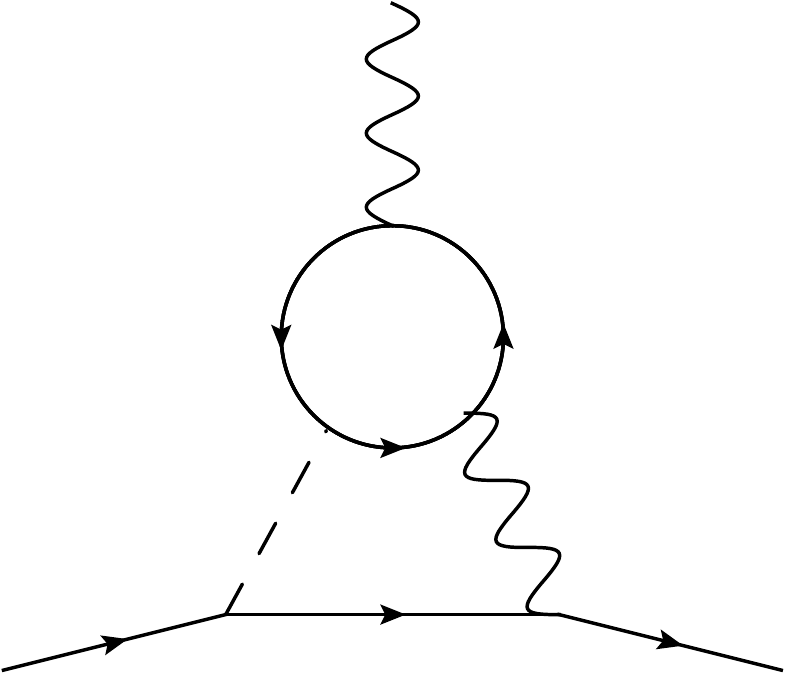}\hspace{2ex}
		\label{fig:BarrZee1}
	}
	\subfloat[]{
		\centering
		\includegraphics[width=4cm]{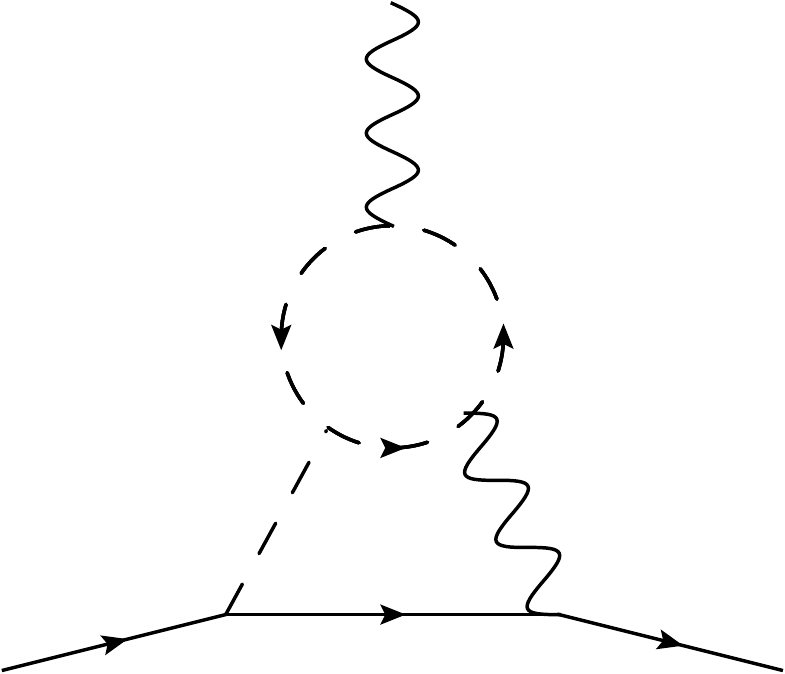}\hspace{2ex}
		\label{fig:BarrZee2}
	}
	\subfloat[]{
		\centering
		\includegraphics[width=4cm]{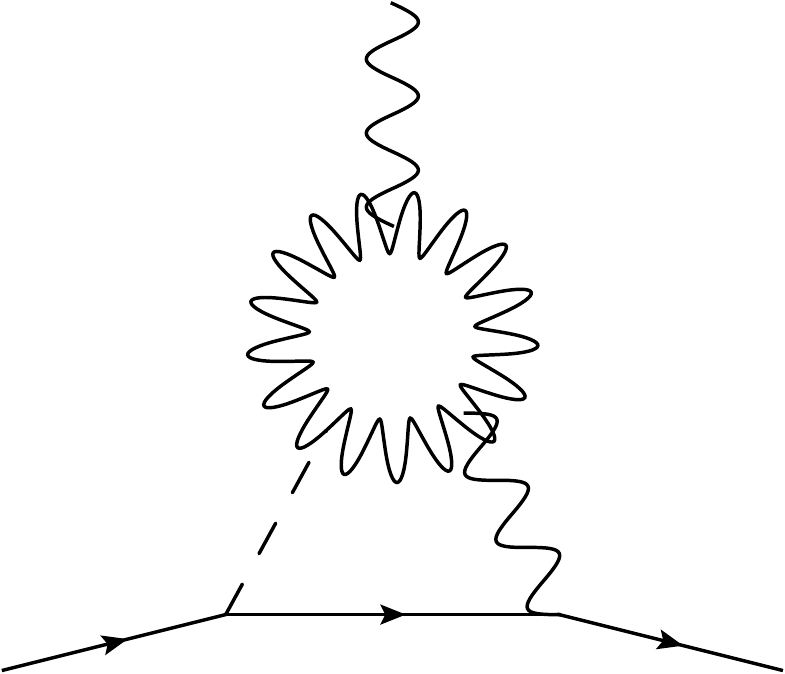}
		\label{fig:BarrZee3}
	}
	
	\vspace{2ex}
	
	\subfloat[]{
		\centering
		\includegraphics[width=4cm]{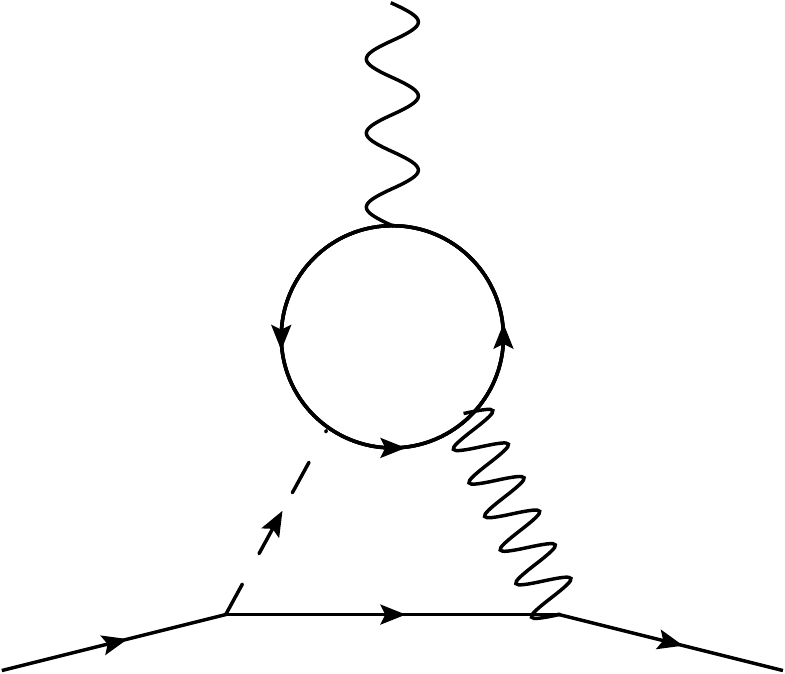}\hspace{2ex}
		\label{fig:BarrZee4}
  }
	\subfloat[]{
		\centering
		\includegraphics[width=4cm]{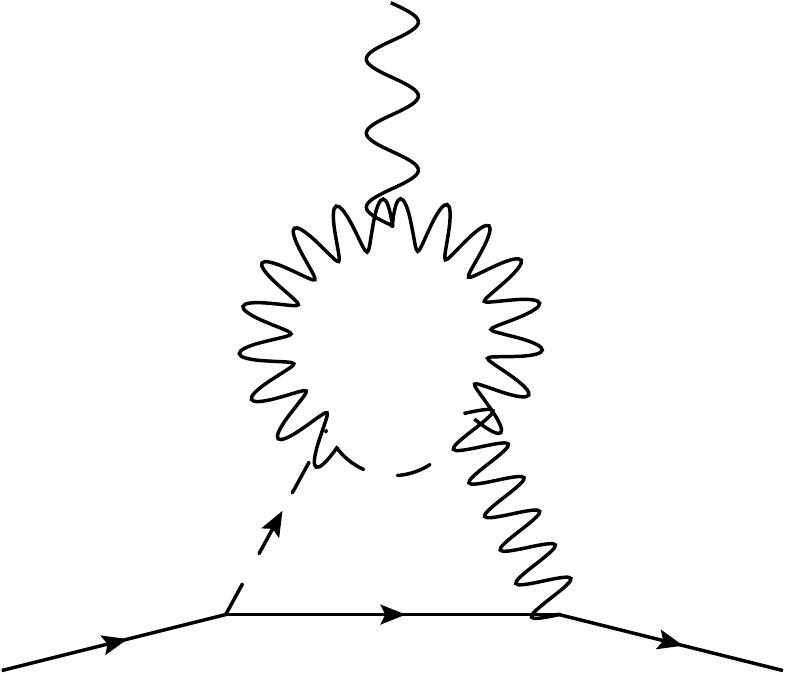}\hspace{2ex}
		\label{fig:BarrZee5}
	}
	\subfloat[]{
		\centering
		\includegraphics[width=4cm]{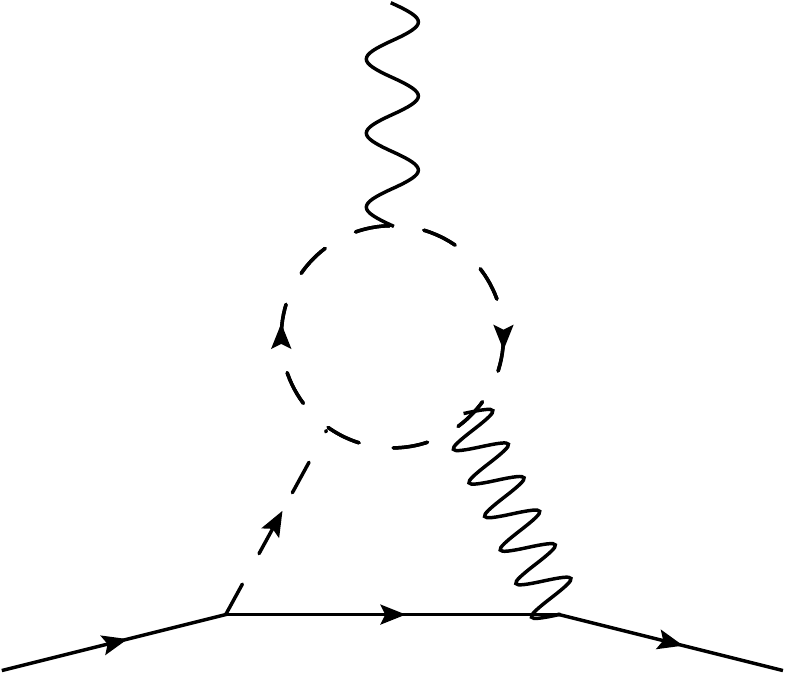}
		\label{fig:BarrZee6}
	}
	\caption{Barr--Zee diagrams relevant for the Wilson coefficients $c_{L,R}^{\ell_f \ell_i}$.}
	\label{fig:BarrZee}
\end{figure*}

It is well known that two-loop Barr--Zee-type diagrams \cite{Barr:1990vd} (see Fig.~\ref{fig:BarrZee}) can dominate in some regions of parameter space due to enhancement factors $m_f^2/m_\mu^2$ from fermions $f$ in the loop, which can overcome the additional loop suppression $\alpha_\mathrm{EM}/\pi$. For the 2HDM-X, there is an additional $\tan\beta$ enhancement for $A$, $H$, and enhanced $2$-loop contributions come from the $\tau$ in the loop. However, there are in addition other important diagrams \cite{Chang:1993kw,Hisano:2006mj,Jung:2013hka,Abe:2013qla,Ilisie:2015tra} that could give relevant contributions to be summarized and converted to our conventions in the following. 

\subsection{Diagram (a)}

The most relevant Barr--Zee diagram contains an additional charged fermion $f$ in the loop (Fig.~\ref{fig:BarrZee1}). 
\begin{align}
	c_R^{\ell_f\ell_i,(a)} = \frac{-e^3}{128 \pi^4 } \sum_{f_j}  \frac{N_f^c Q^2_f}{m_{\ell_i}}  \sum_{i=h,H,A}  \Gamma_{\ell_f\ell_i}^i \Gamma_{f_jf_j}^i  \frac{m^{f_j}}{m_i^2} g_i^{(a)}(r_{f_j}^i) ,
\end{align}
where $N_f^c$ and $Q_f$ are the number of colours and charge of fermion $f$, respectively, and $r_f^i \equiv m_f^2/m_i^2$.
The loop function is given by
\begin{align}
	g_i^{(a)}(r) = \int_0^1 \dd x\, \frac{N_i(x)}{x(1-x) - r} \ln\left( \frac{x(1-x)}{r} \right) ,
\end{align}
where
\begin{align}
	N_h(x) = N_H(x) = 2x(1-x) - 1 \,, \quad N_A(x) = -1 \,.
\end{align}

\subsection{Diagram (b)}

The diagram from Fig.~\ref{fig:BarrZee2} contains a charged scalar in the loop and hence depends on the $H^+ H^- H_i^0$ self-interactions.
The relevant dimensionless trilinear Higgs couplings are given by
\begin{align}
\begin{split}
\lambda_{H^+ H^- h} &=  - \frac{1}{v^2}\left( m_{H}^2-m_h^2\right) \frac{\sin(2\alpha+2\beta) \sin(\alpha-\beta)}{\sin (4\beta)}\\
&\quad - \frac{1}{v^2}\left( m_{H^+}^2-\frac{m_h^2}{2}\right) \sin(\alpha-\beta)\\
&\quad +\Delta \lambda \tan (2\beta) \cos(\alpha+\beta)\,,
\end{split}\\
\begin{split}
\lambda_{H^+ H^- H} &= + \frac{1}{v^2}\left( m_{H}^2-m_h^2\right) \frac{\sin(2\alpha+2\beta) \cos(\alpha-\beta)}{\sin (4\beta)}\\
&\quad + \frac{1}{v^2}\left( m_{H^+}^2-\frac{m_H^2}{2}\right) \cos(\alpha-\beta)\\
&\quad + \Delta \lambda \tan (2\beta) \sin(\alpha+\beta)\,,
\end{split}
\label{eq:H+H-H}
\end{align}
and $\lambda_{H^+ H^- A} = 0 $. $\Delta \lambda \equiv \lambda_2-\lambda_1$ is a free parameter~\cite{Skiba:1992mg}.
While the last two terms in Eq.~\eqref{eq:H+H-H} vanish in the SM-like limit $\sin (\beta-\alpha)=1$, the first is still suppressed by $1/\tan\beta$ for large $\tan\beta$.
The Wilson coefficient is then given by~\cite{Ilisie:2015tra}
\begin{align}
\begin{split}
	c_R^{\ell_f\ell_i,(b)} &= \frac{ e^3}{128\sqrt2 \pi^4} \frac{v}{m_{\ell_i}}  \\
	&\quad\times\sum_{i=h,H,A} \frac{\Gamma_{\ell_f\ell_i}^i}{m_i^2} \zeta^i \lambda_{H^+H^-H_i^0} g_i^{(b)}\left( \frac{m_{H^+}^2}{m_i^2} \right)  ,
	\end{split}
\end{align}
where $\zeta^h = -\zeta^H = -\zeta^A = 1$ and the loop function is
\begin{align}
	g_{h,H,A}^{(b)}(r) = \int_0^1 \dd x\, \frac{x(1-x)}{x(1-x) - r} \ln\left( \frac{r}{x(1-x)} \right) .
\end{align}

\subsection{Diagram (c)}

The contribution of the diagram in Fig.~\ref{fig:BarrZee3} with a $W$ in the loop is~\cite{Ilisie:2015tra}:
\begin{align}
	c_R^{\ell_f\ell_i,(c)} = \frac{G_F e^3}{64 \pi^4} \sum_{i=h,H} \frac{v}{m_{\ell_i}} \Gamma_{\ell_f\ell_i}^i \zeta^i \mathcal{R}_{i1} \, g_i^{(c)}\left( \frac{m_W^2}{m_i^2} \right) ,
\end{align}
where $\mathcal{R}_{h1} = \sin(\beta-\alpha)$, $\mathcal{R}_{H1} = -\cos(\beta-\alpha)$ and the loop function is
\begin{align}
\begin{split}
	g_{h,H}^{(c)}(r) &= \frac{1}{2} \int_0^1 \dd x\, \left[\frac{x(1-x) - x r [3x(4x-1)+10]}{x(1-x) - r}\right.\\
	&\qquad \left. \times \ln\left( \frac{r}{x(1-x)} \right) \right].
\end{split}
\end{align}

\subsection{Diagrams (d), (e), (f)}

Finally, there are three types of Barr-Zee diagrams, where the virtual $H_i^0$ and $\gamma$ propagators are replaced by $H^+$ and $W$: Figs.~\ref{fig:BarrZee4} (where the fermion line is a $t$/$b$- or $b$/$t$-loop), \ref{fig:BarrZee5} and \ref{fig:BarrZee6}. 
Defining the loop function
\begin{align}
	G(r^a, r^b) = \frac{\ln\left( \frac{r^a x + r^b(1-x)}{x(1-x)} \right)}{x(1-x) - r^a x - r^b(1-x)} \,,
\end{align}
and the matrix
\begin{align}
	\mathcal{R} = \left( \begin{matrix} \sin(\beta-\alpha) &\cos(\beta-\alpha) & 0 \\ -\cos(\beta-\alpha) & \sin(\beta-\alpha) & 0 \\ 0 & 0 & 1 \end{matrix} \right) ,
\end{align}
the diagrams in Fig.~\ref{fig:BarrZee4}, \ref{fig:BarrZee5} and~\ref{fig:BarrZee6} are~\cite{Ilisie:2015tra}
\begin{widetext}
\begin{align}
	\begin{split}
		c_R^{\ell_f\ell_i,(d)} &= \frac{- e^3}{1024\pi^4 \sin^2\theta_w } \frac{N^c_t V_{tb}^*}{(m_{H^+}^2 - m_W^2)} \int_0^1 \dd x\, \left( Q_t x + Q_b(1-x) \right) \left[ G\left( \frac{m_t^2}{m_{H^+}^2}, \frac{m_b^2}{m_{H^+}^2} \right) - G\left( \frac{m_t^2}{m_W^2}, \frac{m_b^2}{m_W^2} \right) \right]\\
			&\quad \times \left[ \left( {\Gamma_{tb}^{H^+,LR}}^* {\Gamma_{\nu_f\ell_i}^{H^+}} \right) \frac{m_b}{m_{\ell_i}} x(1-x) - \left( {\Gamma_{tb}^{H^+,RL}}^* {\Gamma_{\nu_f\ell_i}^{H^+}} \right) \frac{m_t}{m_{\ell_i}} x(1+x) \right]  ,
	\end{split}\\
	\begin{split}
		c_R^{\ell_f\ell_i,(e)} &= \frac{G_F e^3}{64\pi^4 \sqrt{2}} \frac{1}{8 \sin^2\theta_w} \sum_i \frac{v}{m_{\ell_i}} {\Gamma_{\nu_f\ell_i}^{H^+}} (\mathcal{R}_{i1} ( \mathcal{R}_{i2} - i \mathcal{R}_{i3} ))^* \int_0^1 \dd x \, x^2 \\
			&\quad \times \left( \frac{(m_{H^+}^2 + m_W^2 - m_i^2) (1-x) - 4 m_W^2}{m_{H^+}^2 - m_W^2} \right) \left[ G\left( \frac{m_W^2}{m_{H^+}^2}, \frac{m_i^2}{m_{H^+}^2} \right) - G\left( 1, \frac{m_i^2}{m_W^2} \right) \right] ,
	\end{split}\\
	\begin{split}
		c_R^{\ell_f\ell_i,(f)} &= \frac{G_F e^3}{64\pi^4 \sqrt{2}} \frac{1}{4 \sin^2\theta_w}\frac{v^2}{(m_{H^+}^2 - m_W^2)} \sum_i \frac{v}{m_{\ell_i}} {\Gamma_{\nu_f\ell_i}^{H^+}} ( \mathcal{R}_{i2} - i \mathcal{R}_{i3} )^* \zeta^i \lambda_{H^+H^-H^0_i} \\
			&\quad \times \int_0^1 \dd x \, x^2(x-1)  \left[ G\left( 1, \frac{m_i^2}{m_{H^+}^2} \right) - G\left( \frac{m_{H^+}^2}{m_W^2}, \frac{m_i^2}{m_W^2} \right) \right] .
	\end{split}
\end{align}
\end{widetext}

\bibliography{AMM-tauonic}% Produces the bibliography via BibTeX.

\end{document}